\documentclass[oneside,lettersize,journal]{IEEEtran}
\usepackage{amsmath,amsfonts}
\usepackage{algorithmic}
\usepackage{algorithm}
\usepackage{array}
\usepackage{textcomp}
\usepackage{stfloats}
\usepackage{url}
\usepackage{verbatim}
\usepackage{graphicx}
\usepackage{cite}
\usepackage{upgreek}
\usepackage{amssymb}
\usepackage{xcolor}
\usepackage{hyperref}
\usepackage{balance}
\ifCLASSOPTIONcompsoc
\usepackage[caption=false, font=normalsize, labelfont=sf, textfont=sf]{subfig}
\else
\usepackage[caption=false, font=footnotesize]{subfig}
\fi

\newtheorem{remark}{Remark}

\begin{document}

\title{Two-Stage Signal Reconstruction  for  Amplitude-Phase-Time Block Modulation-based  Communications}


\author{Meidong~Xia,~\IEEEmembership{Student Member,~IEEE,}
		Min Fan,~\IEEEmembership{Student Member,~IEEE,}
		Wei~Xu,~\IEEEmembership{Fellow,~IEEE,}
		Haiming~Wang,~\IEEEmembership{Member,~IEEE,}
		and Xiaohu~You,~\IEEEmembership{Fellow,~IEEE}%

		\thanks{This work was supported in part by the National Natural Science Foundation of China under Grants 62550015 and 62571112, and in part by the Science Foundation of Jiangsu Province of China under Grant BG2025001. (\emph{Corresponding authors: Wei Xu and Haiming Wang.})}
		\thanks{Meidong  Xia, Wei Xu, and Xiaohu You are with the National Mobile Communications Research Laboratory, Southeast University, Nanjing 211189, China, and also with the Pervasive Communication Research Center, Purple Mountain Laboratories, Nanjing 211111, China (e-mail: \{meidong.xia, wxu, xhyu\}@seu.edu.cn).} 
		\thanks{Min Fan and Haiming Wang are with the State Key Laboratory of Millimeter Waves, Southeast University, Nanjing 211189, China, and also with the Pervasive Communication Research Center, Purple Mountain Laboratories, Nanjing 211111, China (e-mail: \{minfan, hmwang\}@seu.edu.cn).}
		}




\maketitle


\begin{abstract}
Operating power amplifiers (PAs) at lower input back-off (IBO) levels is an effective way to improve PA efficiency, but often introduces severe nonlinear distortion that degrades transmission performance.  Amplitude-phase-time block modulation (APTBM) has recently emerged as an effective solution to this problem. The intrinsic amplitude and phase constraints of each APTBM block can be leveraged to mitigate PA-induced nonlinear distortion via constraint-guided signal reconstruction.
However, existing reconstruction methods apply these constraints only heuristically and statistically, limiting the achievable IBO reduction and PA efficiency improvement. This paper addresses this limitation by decomposing the nonlinear distortion into dominant and residual components, and accordingly develops a novel two-stage signal reconstruction algorithm consisting of  coarse and fine reconstruction stages. The coarse reconstruction stage   eliminates the dominant distortion by jointly exploiting the APTBM block structure and PA nonlinear characteristics. Subsequently, the fine reconstruction stage minimizes the residual distortion by casting it as a nonconvex optimization problem subject to explicit APTBM constraints, for which a  closed-form solution is derived. The proposed algorithm is validated through comprehensive numerical simulations and testbed experiments. Results show that, without compromising transmission quality, the proposed algorithm enables an additional IBO reduction of approximately 5 dB in simulations and 2 dB in experiments over baseline methods, yielding relative PA efficiency improvements of 77.8\% and 30.9\%, respectively.
\end{abstract}


\section{Introduction}
\label{sec:introduction} 

\IEEEPARstart{T}{he} advent of sixth-generation (6G) wireless networks will impose unprecedented  quality-of-service (QoS) requirements, particularly regarding data rate, connectivity, latency, reliability, and coverage \cite{Wang2023OnTheRoad}. Meeting these demands will likely necessitate the utilization of wider bandwidth resources, the deployment of increasingly dense network infrastructures, and the execution of more computationally intensive signal processing tasks \cite{Xu2023EdgeLearning,YangPrivacySecurityTrustworthiness2025}. These developments, however, are anticipated to drive a substantial increase in overall network energy consumption. Recent studies suggest that by 2030, the information and communication technology (ICT) sector could account for more than 20\% of global energy consumption \cite{Zhang2024Toward}, underscoring an urgent need to improve the energy efficiency (EE) of future wireless networks \cite{Yao2025Energy}. 
To align with long-term sustainability goals, it is estimated that 6G networks need to achieve an EE improvement of up to two orders of magnitude compared to current fifth-generation (5G) systems \cite{Wang2023OnTheRoad}.

The primary power consumer in a typical cellular network is the base station (BS), accounting for approximately 50--60\% of the total power usage \cite{Han2011GreenRadio}. Within a BS, the power amplifier (PA) is the most energy-intensive component, responsible for 50--80\% of the total energy consumption \cite{huawei2020, Andersson2022Improving}. {Consequently, maximizing the power-added efficiency (PAE) of PAs, defined as the ratio of the added radio-frequency (RF) power to the consumed direct-current (DC) power, is critical for enhancing system-level EE. Specifically, an improved PAE allows the desired RF output power to be delivered using significantly less DC power, directly reducing overall energy consumption.}
However, a fundamental trade-off exists between PA efficiency and transmission fidelity. While PAs achieve peak PAE near their saturation region, operating in this regime induces severe nonlinear distortion that degrades signal quality and communication performance \cite{Auer2011HowMuchEnergy, He2024Aunified}. To circumvent this distortion and preserve signal integrity, conventional wireless systems operate PAs with a large input back-off (IBO), keeping them well below saturation \cite{Auer2011HowMuchEnergy}. Although effective for linearity, this conservative approach is inherently  inefficient: it forces the use of higher-rated PAs operating in a low-PAE region to achieve the required output power, ultimately increasing both energy consumption and hardware cost \cite{Abdelaziz2018Digital}.

To address this limitation, extensive techniques  have been developed  to enable PA operation closer to saturation, i.e., with a lower IBO, while preserving  transmission performance. Among these,  digital pre-distortion (DPD)   is widely adopted in practical systems  \cite{guan2014green, Katz2016Evolution}. DPD    compensates for PA nonlinearity by applying an inverse distortion to the  input signal, thereby linearizing the overall PA response \cite{Zhao2023Alowcomplexity}.  However, a fundamental limitation of DPD is that it can only linearize the PA response up to its saturation point. Consequently, a relatively high IBO remains necessary, particularly when transmitting signals with a high peak-to-average power ratio (PAPR) \cite{Feys2025TowardEnergyEfficient}. Furthermore, DPD implementations typically necessitate significant hardware overhead, including high-speed analog-to-digital converters (ADCs) and dedicated feedback loops. This dramatically increases system complexity and cost, especially in high-frequency scenarios where hardware is highly expensive and design margins are inherently constrained \cite{xuNewPathIntegrated2026, He2024Unlocking}.

To circumvent the hardware overhead associated with DPD, distortion-aware signal processing approaches have been developed.
Rather than attempting to linearize the PA response, these methods explicitly incorporate PA nonlinearity into the algorithmic design to enable low-IBO operation. 
For instance, the authors in \cite{Liu2020Optimal} modeled PA nonlinearity using the Saleh model \cite{Saleh1981Frequency} and integrated this model into beamforming optimization to maximize the achievable rate under nonlinear distortion. To mitigate the analytical complexity of the Saleh model, a more tractable polynomial approximation was adopted in \cite{Aghdam2019Distortionaware}, which leveraged Bussgang decomposition \cite{Bussgang1952Crosscorrelation} to facilitate distortion-aware beamforming. Similarly, joint channel estimation and signal detection algorithms utilizing polynomial-based PA models were proposed in \cite{Priya2023Channel} and \cite{Jian2025Sparse}. While these approaches can achieve considerable performance gains in the presence of nonlinear distortion, they heavily rely on specific and  intricate PA models. Consequently, even when employing simplified polynomial approximations and Bussgang-based linearization, the resulting algorithms remain computationally intensive and analytically challenging.

Furthermore, the detrimental effects of nonlinear distortion become more pronounced when transmitted signals exhibit a high PAPR, as is common in orthogonal frequency-division multiplexing (OFDM) systems. To address this vulnerability, various PAPR reduction techniques have been developed.
For example, the iterative clipping and filtering (ICF) method proposed in \cite{Wang2011Optimized} alternately clips the signal to constrain peak amplitudes and applies filtering to suppress out-of-band emissions. In another approach, the authors in \cite{Chen2024Constant} integrated a phase modulator into the OFDM transmitter to embed the information waveform directly into the signal phase. This   technique generates a constant-envelope signal, albeit at the expense of bandwidth expansion. Alternatively, the PAPR reduction task was formulated as a mathematical optimization problem in \cite{Guo2016Transmitter}. By minimizing the peak power of the transmitted signal, the resulting PAPR was proven to be upper-bounded by a function inversely proportional to the number of RF chains. Although these methods effectively reduce PAPR and facilitate low-IBO operation without requiring additional hardware or specific PA models, they often incur a significant loss in spectral efficiency (SE) \cite{Moghadam2018OnEnergy}.

To overcome the limitations of the aforementioned techniques,
amplitude-phase-time block modulation (APTBM) has recently been proposed  \cite{fanAmplitudephasetimeBlockModulation2023}. APTBM encodes information in blocks, each containing two symbols subject to inherent amplitude and phase constraints. These constraints enable  constraint-guided signal reconstruction at the receiver, mitigating nonlinear distortion in a hardware-free and model-agnostic manner while maintaining a favorable trade-off between SE and EE \cite{fanAmplitudephasetimeBlockModulation2023}. 
Naturally, the overall  performance of APTBM-based systems  depends crucially on the effectiveness of this signal reconstruction. However, the existing method in \cite{fanAmplitudephasetimeBlockModulation2023}  applies the amplitude and phase constraints only heuristically and statistically, rather than explicitly integrating them  into the reconstruction process. As a result, it achieves only partial distortion mitigation, leaving considerable residual distortion that degrades system  performance.  {Consequently, the PA still needs to operate with a relatively high IBO to ensure transmission quality, which severely undermines the potential PAE gains offered by APTBM.} This limitation highlights the need for advanced reconstruction techniques that explicitly exploit the APTBM constraints to enable further IBO reduction.

Three key challenges arise in this regard. First, the APTBM constraints are highly nontrivial, which complicates their effective integration into the reconstruction process. Second, although nonlinear distortion exhibits common structural patterns across different PAs, how to effectively leverage these patterns to enhance reconstruction performance remains a critical open challenge. Third, the periodicity of phase angles leads to intrinsic ambiguities in phase manipulation, posing additional challenges for achieving exact phase recovery.  Motivated by these challenges, we propose a novel two-stage signal reconstruction framework that progressively mitigates nonlinear distortion in APTBM-based systems. By enabling operation at a reduced IBO, our proposed framework significantly enhances PA efficiency.
The main contributions of this paper are summarized as follows:
\begin{itemize}
\item We decompose the PA-induced nonlinear distortion into a dominant component and a residual component, and accordingly develop a   two-stage reconstruction framework comprising coarse and fine reconstruction stages. 
\item In the coarse reconstruction stage, we design distortion compensation schemes that leverage both the APTBM block structure and PA nonlinear characteristics to eliminate the dominant distortion. 
\item In the fine reconstruction stage, we minimize the residual distortion by formulating a nonconvex optimization problem subject to explicit APTBM constraints. By exploiting the inherent geometric symmetry of the APTBM blocks, we first unambiguously estimate the initial phase. Based on this estimate, we subsequently derive a closed-form solution to the reduced optimization problem,   thereby yielding the fully reconstructed APTBM blocks.
\item The proposed  algorithm is evaluated through numerical simulations and testbed experiments conducted in both sub-6 GHz and millimeter-wave (mmWave) bands.  Extensive results demonstrate that the proposed method significantly reduces the required IBO and achieves substantial PAE gains compared to existing methods.
\end{itemize}

The remainder of this paper is organized as follows.
Section~\ref{sec:system_model} describes the system model for APTBM-based nonlinear transmissions.
Section~\ref{sec:proposed_SR} presents the proposed   signal reconstruction algorithm in detail.
Performance evaluations through   numerical simulations and testbed experiments are provided in Section~\ref{sec:simulation}. Finally, Section~\ref{sec:conclusions} concludes the paper and discusses future research directions.

\subsubsection*{Notations}
In this paper, $(\cdot)^T$, $(\cdot)^*$,
and $(\cdot)^H$ denote the transpose, complex  conjugate,
and conjugate transpose, respectively. $\mathrm{Re}(\cdot)$ and $\mathrm{Im}(\cdot)$ represent the real and imaginary parts of a complex number. The operators $\sqrt{\cdot}$, $e^{(\cdot)}$,  $\log_{a}(\cdot)$, $\left|\cdot\right|$, $\left\|\cdot\right\|$, $\angle(\cdot)$, $\sin(\cdot)$, $\cos(\cdot)$, $\tan(\cdot)$, and $\mathrm{atan2}(\cdot,\cdot)$ denote the square root, exponential,   base-$a$ logarithm, complex modulus, Euclidean
norm, phase angle, sine, cosine, tangent, and four-quadrant arctangent functions, respectively. The imaginary unit is denoted by $\jmath$. The notation $\left\{x_i\right\}$ indicates a set indexed by $i$. $\mathbb{R}$ and $\mathbb{C}$ denote the real and complex number sets, and $\mathcal{O}$ represents big-O notation. The symbols $\circledast$, $\odot$, $\mathbb{E}$, $\triangleq$, $\subset$, and $\in$ denote convolution, Hadamard product, expectation, definition, subset relation,
and set membership, respectively. 
Finally, $f(x)\big|_{x=a}$ denotes the evaluation of $f(x)$ at $x=a$.

\section{System Model for APTBM-based Nonlinear Transmissions} 
\label{sec:system_model}
This section first reviews the fundamentals of the APTBM scheme and then formulates the corresponding nonlinear transmission system model.

\begin{figure*}[t]
	\centering
	\subfloat[]{\includegraphics[width=1.85in]{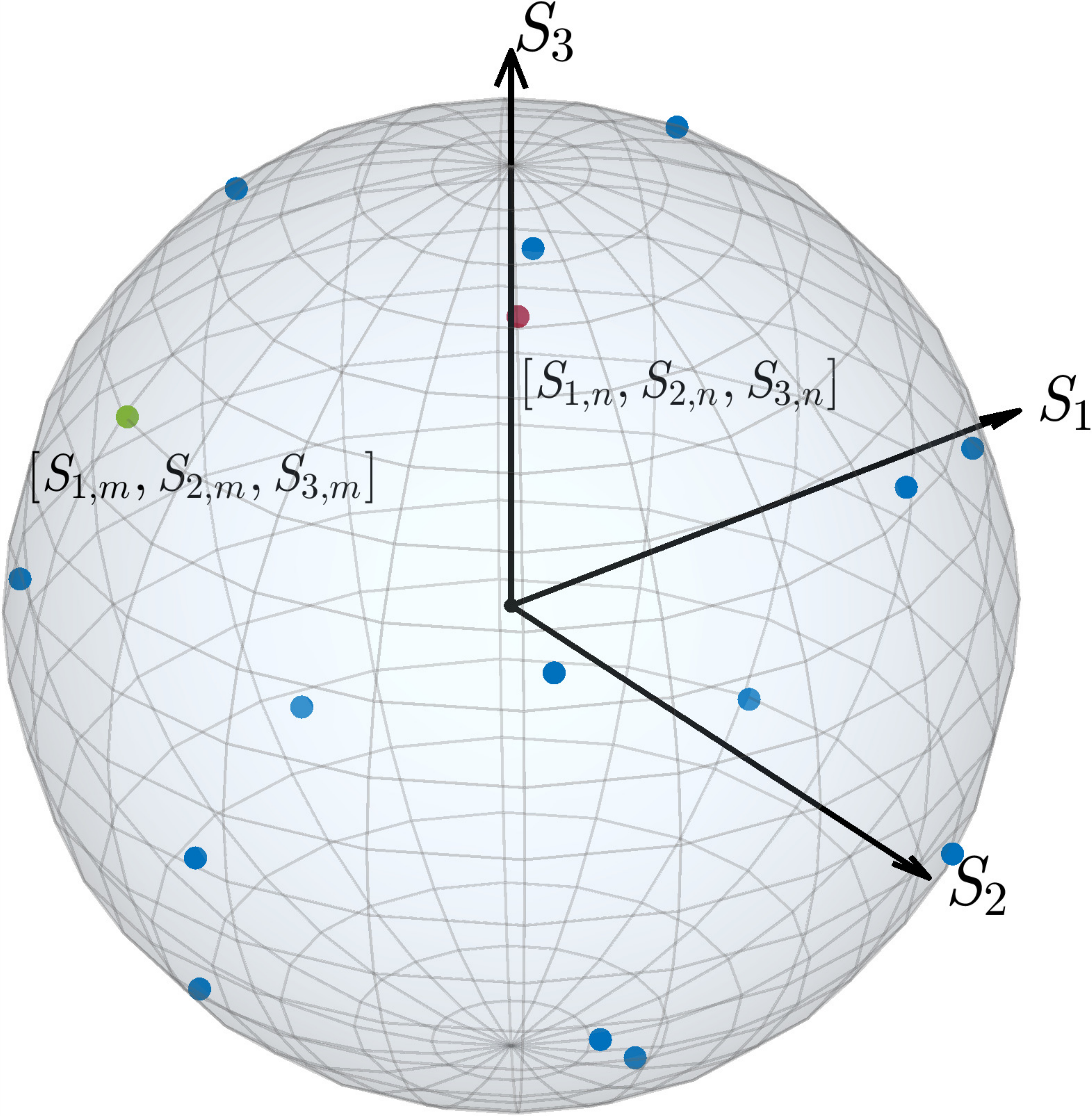} \label{sphere}}%
	\hfill
	\centering
	\subfloat[]{\includegraphics[width=2in]{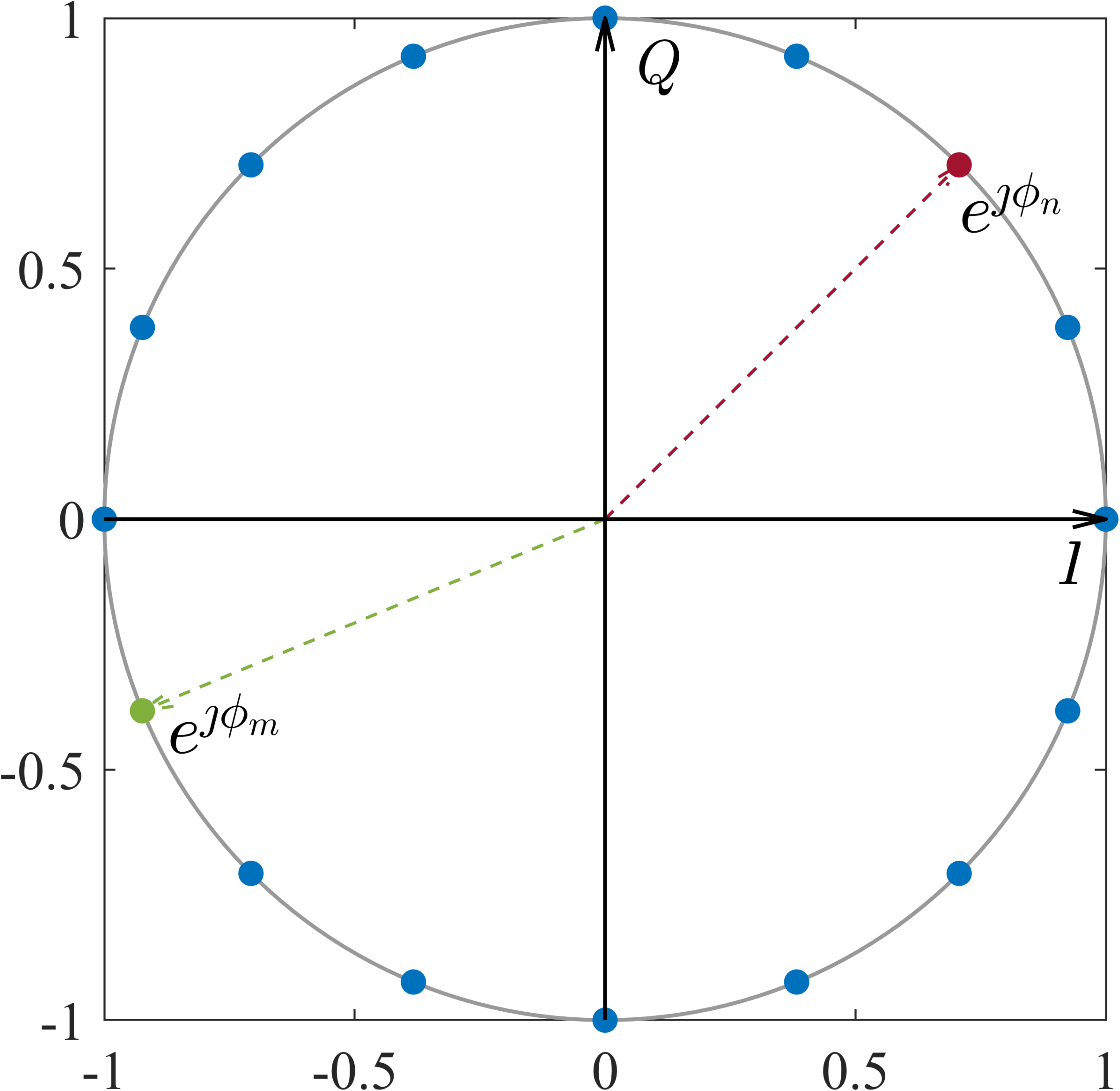} \label{phase}}%
	\hfill
	\centering
	\subfloat[]{\includegraphics[width=2in]{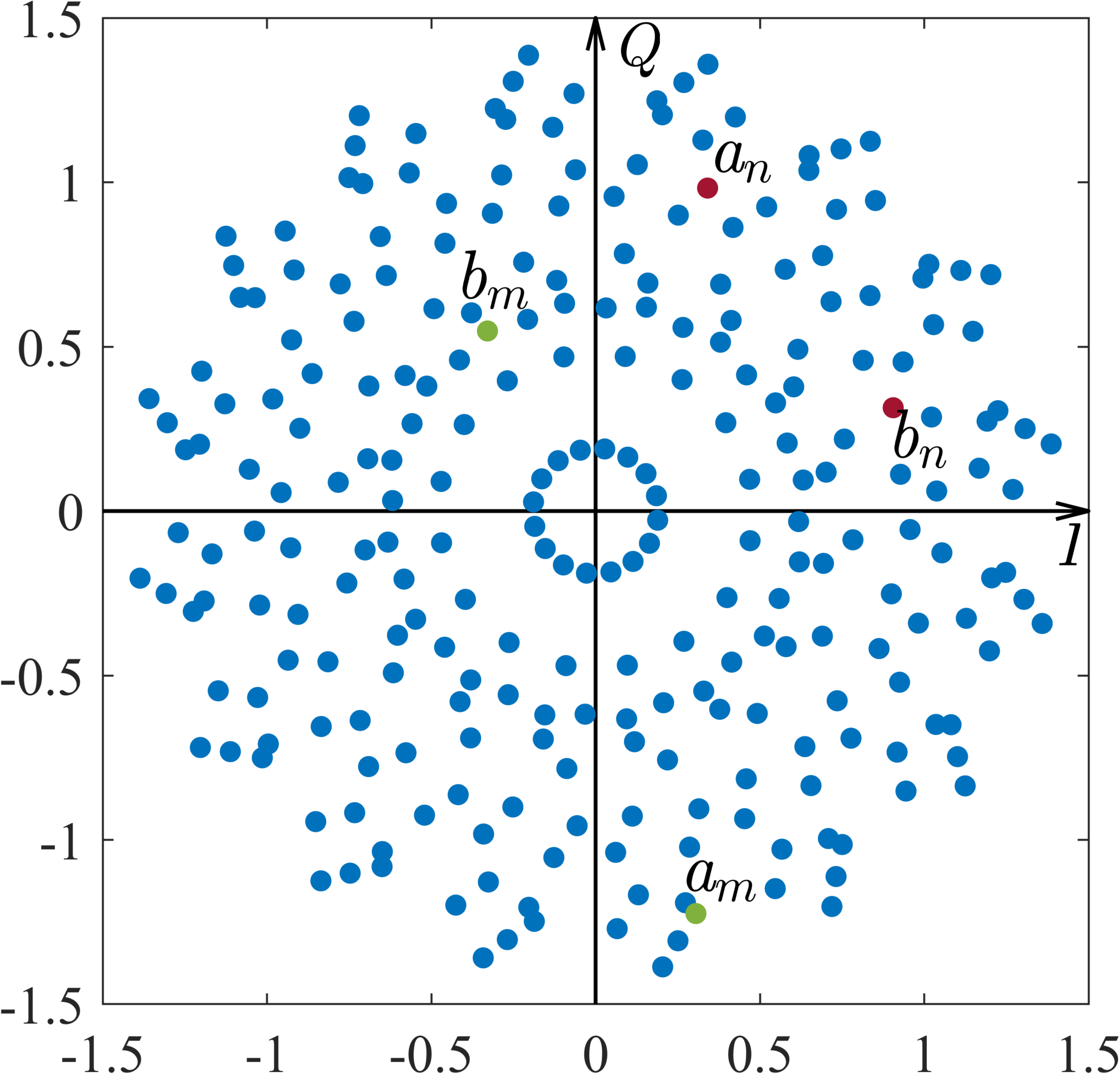} \label{aptbm1}}%
	\caption{Constellation representation of the APTBM blocks. (a) Spherical surface alphabet $\mathbb{S}_L$ with $L=16$. (b) Initial phase alphabet $\mathbb{S}_M$ with $M=16$. (c) APTBM symbol alphabet $\mathbb{S}_{ML}$ with $M=16$ and $L=16$. Two representative blocks are highlighted in distinct colors to exemplify the mapping procedure.}
	\label{fig:APTBM_circular}
\end{figure*}

\subsection{Review of APTBM}\label{subsec:aptbm}

APTBM is a recently developed block modulation scheme that encodes information bits onto the surface of a Poincaré sphere and leverages the initial phase as an additional degree of freedom   to improve SE \cite{fanAmplitudephasetimeBlockModulation2023}.

Let $\left\{\mathbf{c}_n\right\}$, $n  \in \left\{1, \cdots, N\right\}$, 
denote the   APTBM blocks, where  $N$ is the total number of  blocks.
The $n$-th   block is  expressed as
\begin{equation}
\mathbf{c}_n = \left[ a_n,   b_n \right]^T \in \mathbb{C}^2, \ \forall n,  \label{eq:aptbm_block}
\end{equation}
consisting of two consecutive time-domain symbols $a_n$ and $b_n$.
The APTBM  blocks are generated as follows \cite{fanAmplitudephasetimeBlockModulation2023} 
\begin{subequations}
	\begin{align}
		a_n &= e^{\jmath \phi_n } 
			\sqrt{\frac{  P + S_{1, n} }{2}} 
				e^{-\jmath {\theta_n}}, \ \forall n, \label{eq:aptbm_a}\\
		b_n &= e^{\jmath \phi_n }
			\sqrt{\frac{ P - S_{1, n} }{2}} 
				e^{\jmath {\theta_n}}, \ \forall n, \label{eq:aptbm_b}
	\end{align}
	\label{eq:aptbm_block_generate}%
\end{subequations}
where $P$ denotes the total   power per block, and $\theta_n = \frac{1}{2} \mathrm{atan2}  \Big(S_{3,n}, S_{2,n}\Big)$. Here,  $\phi_n$ represents the initial phase of the $n$-th  block, which is uniformly distributed over $(-\pi,   \pi]$.
In addition, the vector $\left[ S_{1,n},   S_{2,n},   S_{3,n} \right]^T$ denotes the  parameter vector associated with the $n$-th block, representing the coordinates of a state  point on the  surface of a   Poincaré sphere with 
radius $P$ \cite{henarejos3DPolarizedModulation2018}.

The APTBM symbol alphabet is formulated using $L$ distinct state points on the surface of a Poincaré sphere combined with $M$ distinct initial phases. Let $\mathbb{S}_{ML} \subset \mathbb{C}^2$ denote this composite alphabet, such that the $n$-th transmitted block satisfies $\mathbf{c}_n \in \mathbb{S}_{ML}$. This alphabet is  defined by two independent components:
\begin{itemize}
	\item A spherical surface alphabet, $\mathbb{S}_{L} \subset \mathbb{R}^3$, such that $\left[S_{1,n}, S_{2,n}, S_{3,n}\right]^T \in \mathbb{S}_{L}$ for all $n$.
	\item An initial phase alphabet, $\mathbb{S}_{M} \subset \mathbb{R}$, such that the initial phase of the $n$-th block satisfies $\phi_n \in \mathbb{S}_{M}$.
\end{itemize}
Under this modulation scheme, $\log_2 M$ bits are mapped onto an initial phase, and $\log_2 L$ bits are mapped onto  a spherical state point. Therefore, each APTBM block carries a total of $\log_2{ML}$ information bits.
To visualize this, Fig.~\ref{fig:APTBM_circular} depicts the constellation structure for $M=16$ and $L=16$, where two representative blocks are highlighted in distinct colors to exemplify the mapping procedure.

The key feature of APTBM is that the two consecutive time-domain symbols within each block satisfy the following amplitude and phase constraints \cite{fanAmplitudephasetimeBlockModulation2023}
\begin{subequations}
\begin{align}
	\left| a_n \right|^2 +  \left| b_n \right|^2 & = P, 
		\ \forall n,   \label{eq:aptbm_power} \\
	\angle  \left( a_n \right)   +  \angle  \left( b_n \right)   & = 2 \phi_n,
		\ \forall n. \label{eq:aptbm_phase}
\end{align} \label{eq:aptbm_constraints}%
\end{subequations}%
Specifically, the amplitude constraint in \eqref{eq:aptbm_power} guarantees a constant aggregate power $P$ for each block. Concurrently, the phase constraint in \eqref{eq:aptbm_phase} dictates that the sum of the phases of the two symbols equals  twice the initial phase $\phi_n$. These inherent structural constraints serve as the cornerstone of the constraint-guided signal reconstruction strategy developed in this paper.

\subsection{System Model}\label{subsec:system_model}
\begin{figure*}[t]
\centering
\includegraphics[width=0.85\linewidth]{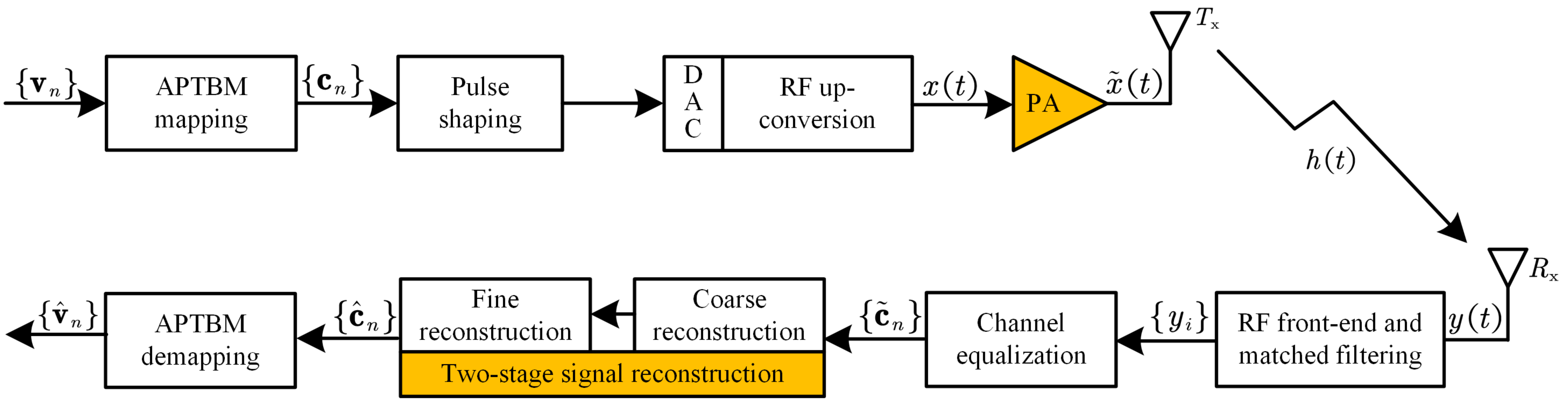}
\caption{System model for APTBM-based  nonlinear transmissions.}
\label{fig:system_diagram}
\end{figure*}

The system model for APTBM-based nonlinear transmission is illustrated in Fig.~\ref{fig:system_diagram}. In the following, a detailed description of each component is provided.

The $n$-th input bit vector, $\mathbf{v}_n = \left[v_{n,1}, \cdots, v_{n,K}\right]^T$, is first mapped to an APTBM  block $\mathbf{c}_n$, where $K = \log_2{ML}$ denotes the number of bits per block, according to
\begin{equation}
\mathbf{c}_n = \mathcal{F} \left( \mathbf{v}_n \right), \quad \forall n.
\label{eq:encoder}
\end{equation}
Here, $\mathcal{F}(\cdot)$ represents the APTBM mapping function, which  maps the $K$ bits onto an initial phase and a spherical surface state point, and subsequently constructs the complex-valued symbol block according to \eqref{eq:aptbm_block_generate}.
Next, the modulated symbols are passed through a pulse shaping filter 
and then processed by the RF front-end, which includes 
digital-to-analog conversion (DAC), RF up-conversion, and power amplification.
Prior to power amplification, the RF signal can be expressed as
\begin{equation}
x(t) =  e^{\jmath 2 \pi f_\mathrm{c} t} \sum_{n=1}^{N} \left( a_n   p\left(t - (2n-1) T_{\mathrm{s}}\right) + b_n   p\left(t - 2n T_{\mathrm{s}}\right) \right), \label{eq:rf_signal}
\end{equation}
where $p(t)$ denotes the pulse shaping filter, $T_{\mathrm{s}}$ is the symbol duration, 
and $f_\mathrm{c}$ represents the carrier frequency.

The RF signal is subsequently amplified by a PA to enhance signal strength and ensure reliable transmission coverage.
For a specific PA, the PAE is defined as \cite{fanAmplitudephasetimeBlockModulation2023}
\begin{equation}
\eta \triangleq \frac{P_{\mathrm{out}}-P_{\mathrm{in}}}{P_{\mathrm{DC}}}, \label{eq:PAE1}
\end{equation}
where $P_{\mathrm{out}}$ is the average PA output power, $P_{\mathrm{in}}$ is the average PA input power, and $P_{\mathrm{DC}}$ is the DC power consumption. For a Class A PA, this can be approximately simplified to  \cite{He2011PowerConsumption}
\begin{equation}
\eta \approx \eta_\mathrm{max} \frac{P_{\mathrm{out}}}{P_{\mathrm{max}}}, \label{eq:PAE}
\end{equation}
where $\eta_\mathrm{max}$ is the maximum achievable PAE (50\% for a Class A PA) and $P_{\mathrm{max}}$ is the maximum output power. Furthermore, the IBO is defined as
\begin{equation}
\mathrm{IBO} \triangleq 10 \log_{10} \left( \frac{P_\mathrm{sat}}{P_\mathrm{in}} \right), \label{eq:IBO}
\end{equation}
where $P_\mathrm{sat}$ represents the input saturation power of the PA.  
Here, $P_{\mathrm{max}}$ and $P_\mathrm{sat}$ are fundamental hardware constraints that remain constant during operation.
Reducing the IBO (i.e., increasing $P_\mathrm{in}$) correspondingly increases $P_{\mathrm{out}}$, provided $P_{\mathrm{max}}$ is not exceeded, which directly improves the PAE according to \eqref{eq:PAE}.  Consequently, to maintain high PAE, it is desirable to operate the  PA at a lower IBO level, albeit at the inevitable expense of exacerbated nonlinear distortion.

The nonlinear characteristics of the PA are represented by a memoryless nonlinear function $\mathcal{T}(\cdot)$, such that the transmitted signal is given by
\begin{equation}
\tilde{x}(t) = \mathcal{T}\left( x(t) \right). \label{eq:transmitted_signal}
\end{equation}
In addition to PA nonlinearity, practical systems are subject to various hardware impairments, such as phase noise, in-phase and quadrature-phase (IQ) imbalance, and quantization noise. However, PA nonlinearity is typically the dominant source of distortion \cite{Moghadam2018OnEnergy}. For clarity, this study focuses exclusively on PA-induced nonlinearity, assuming that other impairments are either negligible or have been sufficiently mitigated by existing techniques.

Upon transmission, $\tilde{x}(t)$ undergoes wireless propagation through a channel with impulse response $h(t)$. The received RF signal is modeled as
\begin{equation}
y(t) = \tilde{x}(t) \circledast h(t) + w(t), \label{eq:received_signal}
\end{equation}
where $w(t)$ denotes additive white Gaussian noise (AWGN).
At the receiver, $y(t)$ is first processed by the RF front-end, which includes RF down-conversion and ADC.
The signal is then passed through a matched filter and downsampled to the symbol rate, yielding the discrete-time sequence $\left\{y_i\right\}$, $i \in \left\{ 1, \cdots, 2N \right\}$, expressed as
\begin{equation}
y_i   = \left( e^{- \jmath 2 \pi f_\mathrm{c} t} y(t) \circledast p^*(-t) \right) 
\big|_{t=iT_{\mathrm{s}}}, \ \forall i.
\label{eq:output_signal}
\end{equation}
To mitigate channel effects, this sequence is subsequently processed by a channel equalizer, 
modeled as a linear filter $\{f_q\}$ of length $Q$. 
The post-equalization signal is given by
\begin{equation}
\tilde{y}_{i} =  \sum_{q=1}^{Q} f_q y_{i-q}, \ \forall i.
\label{eq:output_signal_equalization}
\end{equation}
For block-wise processing, the equalized signal is represented as
$\tilde{\mathbf{c}}_n = \left[ \tilde{y}_{2n-1}, \tilde{y}_{2n} \right]^T$, $\forall n$.

In linear transmission systems, when channel equalization is sufficiently accurate, the post-equalization blocks $\{\tilde{\mathbf{c}}_n\}$ closely approximate the transmitted blocks $\{\mathbf{c}_n\}$, with distortions arising mainly from noise and residual channel effects.
In nonlinear transmission systems, however, even under ideal channel equalization and in the absence of noise,
PA-induced distortion can cause significant deviations between  $\{\tilde{\mathbf{c}}_n\}$ and $\{\mathbf{c}_n\}$.
As a result, an effective distortion mitigation technique, such as the signal reconstruction proposed herein, is essential prior to demapping.
This signal reconstruction process, which exploits the structural constraints of APTBM blocks as defined in \eqref{eq:aptbm_constraints}, constitutes the core focus of this study. 

We denote the symbol blocks after signal reconstruction as $\{\hat{\mathbf{c}}_n\}$, which are intended to closely approximate the original transmitted blocks $\{\mathbf{c}_n\}$. Finally, the reconstructed blocks $\{\hat{\mathbf{c}}_n\}$ are fed into a demodulator, modeled by the inverse mapping function $\mathcal{F}^{-1}(\cdot)$, to obtain the estimated  bit vectors
\begin{equation}
\hat{\mathbf{v}}_n = \mathcal{F}^{-1} \big( \hat{\mathbf{c}}_n \big), \quad \forall n.
\label{eq:decoder}
\end{equation}

It is worth noting that the exact mathematical form of the nonlinear function $\mathcal{T}(\cdot)$ in \eqref{eq:transmitted_signal} does not need to be specified. Because the proposed method is inherently model-agnostic, it can be  applied to various PA models without requiring explicit knowledge of $\mathcal{T}(\cdot)$. In the following section, we detail the proposed two-stage signal reconstruction algorithm.

\section{Proposed Two-Stage Signal Reconstruction Algorithm} 
\label{sec:proposed_SR}

In this section, we first analyze the nonlinear distortion characteristics using a classical PA model. Building upon the obtained insights, we decompose the overall nonlinear distortion into two distinct components and accordingly develop a two-stage signal reconstruction algorithm.

\subsection{Nonlinear Distortion  Decomposition}
In general, the PA-induced nonlinear distortion  manifests in two primary forms {\cite{Saleh1981Frequency,3GPP_R4_163314}:}
\begin{itemize}
\item \textbf{Amplitude-modulation to amplitude-modulation (AM-AM) distortion:} The amplitude of the PA output signal is a nonlinear function of the input signal's amplitude.
\item \textbf{Amplitude-modulation to phase-modulation (AM-PM) distortion:} The phase of the PA output signal is a nonlinear function of the input signal's amplitude.
\end{itemize}
Various mathematical models, such as the Saleh model~\cite{Saleh1981Frequency} and the modified Rapp model~\cite{3GPP_R4_163314}, have been proposed to characterize the AM-AM and AM-PM behaviors of practical PAs. Although these models exhibit distinct nonlinear profiles, they share common structural properties. Here, we adopt the well-known modified Rapp model as an illustrative example, where the AM-AM and AM-PM characteristics are given by
\begin{subequations}
\begin{align}
\left|\tilde{x}(t)\right| & =\frac{g_0\left|x(t)\right|}
		{\left[1+\left(\frac{g_0\left|x(t)\right|}{A_{\text {sat }}}\right)^{2 q_0}\right]^{\frac{1}{2 q_0}}}, \\
\tilde{\psi}(t) & =\frac{\alpha_0\left|x(t)\right|^{q_1}}
		{1+\left(\frac{\left|x(t)\right|}{\beta_0}\right)^{q_2}}+\psi(t).
\end{align} \label{eq:modified_rapp}%
\end{subequations}
Here, $\psi(t)$ and $\tilde{\psi}(t)$ are the phases of the input and output signals, respectively,
and $g_0$, $A_{\text{sat}}$, $\alpha_0$, $\beta_0$, $q_0$, $q_1$, and $q_2$ are the model fitting parameters.


Although the proposed algorithm is model-agnostic, the modified Rapp model offers valuable insights into the nature of nonlinear distortion and provides useful guidance for our algorithm design. 
Specifically, the following observations can be drawn from \eqref{eq:modified_rapp}.

\begin{remark}
\label{remark1}
The output phase consists of the original input phase plus an additional phase shift that depends on the input amplitude. This suggests that the phase shift can be approximately compensated by applying a phase correction based on the average input amplitude.
\end{remark}

\begin{remark}
	\label{remark2}
	When the input amplitude is small, the output amplitude varies almost linearly with the input amplitude. As the input amplitude increases, the output amplitude gradually saturates toward a maximum value, leading to severe nonlinear distortion. This implies that the low-amplitude signals within each APTBM block are more reliable and can be exploited to coarsely compensate for the high-amplitude signals.
\end{remark}

Based on these  insights, we decompose the overall nonlinear distortion into two components: a major distortion component that can be compensated heuristically, and a residual distortion component  requiring further mitigation.  Mathematically, the  distorted signal blocks $\left\{\tilde{\mathbf{c}}_n\right\}$ can therefore be expressed as
\begin{equation}
\tilde{\mathbf{c}}_n = {\mathbf{c}}_n + \mathbf{s}_{n} + \mathbf{r}_n, \ \forall n, \label{eq:signal_reconstruction}
\end{equation}
where $\mathbf{s}_{n}$ denotes the heuristically compensable distortion component, 
and $\mathbf{r}_n$ represents the residual distortion component of the $n$-th APTBM block.

Accordingly, we propose a two-stage reconstruction process that mitigates the overall nonlinear distortion in a progressive manner. The first stage performs coarse reconstruction to eliminate the heuristically compensable distortion $\mathbf{s}_n$, while the second stage conducts fine reconstruction to suppress the remaining residual distortion $\mathbf{r}_n$. The following subsections present each stage in detail.


\subsection{Coarse Reconstruction Stage}
\label{sec:coarse_reconstruction}
The coarse reconstruction stage consists of two key operations: phase compensation and amplitude reconstruction.

Guided by Remark~\ref{remark1}, phase compensation is applied to each received APTBM block to preliminarily correct the PA-induced phase shift. Prior to transmission, the AM-PM characteristics of the PA are either measured offline or extracted from the manufacturer's datasheets. The resulting phase shift  $\phi_{\mathrm{a}}(\cdot)$  is then stored in a lookup table (LUT) as a function of the PA input power. During reception, the compensation angle is retrieved from this LUT using the average PA input power as the index to correct the phase of each block. After applying this compensation, the phase vector for the $n$-th block is updated as
\begin{equation}
	\boldsymbol{\psi}_n =  \angle \left(\tilde{\mathbf{c}}_n \right) - \phi_{\mathrm{a}}\left(\mathbb{E} \left\{  \left| x(t)  \right|^2    \right\}  \right). \label{eq:phase_correction}
\end{equation}
Although this compensation relies on a statistical average and may not match the exact phase shift of every individual symbol, it effectively mitigates the dominant phase shift and confines the residual phase error to a small range. 
It is also worth noting that the AM-PM characteristics of a PA can be influenced by factors such as temperature variations and power supply fluctuations \cite{Zhao2025Broadhand}, which may cause deviations in the compensation angle even under identical operating conditions. Nevertheless, such deviations have a limited impact at this stage, as only coarse phase compensation is required.

Inspired by Remark~\ref{remark2}, amplitude reconstruction coarsely recovers the amplitude of each block by exploiting the less distorted low-amplitude signals alongside the power constraint in \eqref{eq:aptbm_power}.
Following a similar approach to that in \cite{fanAmplitudephasetimeBlockModulation2023}, we reconstruct the amplitude of each block as follows.
Firstly, the amplitude reconstruction coefficient $\xi_n$ is  defined as 
\begin{equation}
	\xi_n  = \frac{1}{1+e^{ \tan \left(0.5 \pi P_{\mathrm{d}, n}\right)}}, 
			\ \forall n, \label{eq:xi}
\end{equation}
where $P_{\mathrm{d}, n}$ denotes the normalized power difference between the two symbols within the $n$-th  symbol block, given by
\begin{equation}
	P_{\mathrm{d}, n}  =\frac{\left|\tilde{a}_{n}\right|^2-
			\left|\tilde{b}_{n}\right|^2}{\left|\tilde{a}_{n}\right|^2+
			\left|\tilde{b}_{n}\right|^2}, \ \forall n. \label{eq:pd}
\end{equation}
It can be observed that when the power of $\tilde{a}_n$ dominates that of $\tilde{b}_n$,
$P_{\mathrm{d}, n}$ approaches one, which in turn drives $\xi_n$ toward zero.
Conversely, when the power of $\tilde{b}_n$ dominates that of $\tilde{a}_n$,
$P_{\mathrm{d}, n}$ approaches minus one, causing $\xi_n$ to approach one.
Leveraging this dynamic behavior, we then reconstruct the amplitude of the $n$-th block as
\begin{equation}
	\breve{\mathbf{c}}_n =\left[\begin{array}{cc}
		\left|\tilde{a}_{n}\right| & \sqrt{P-\left|\tilde{b}_{n}\right|^2} \\
		\sqrt{P-\left|\tilde{a}_{ n}\right|^2} & \left|\tilde{b}_{n}\right|
		\end{array}\right] \left[\begin{array}{cc}
			\xi_n \\ 1-\xi_n
			\end{array}\right], \ \forall n. \label{eq:amplitude_reconstruction}
\end{equation}
The fundamental rationale behind \eqref{eq:amplitude_reconstruction} is that the lower-amplitude symbol serves as a more reliable reference. Consequently, it is assigned a higher weight in the weighted combination of the directly received amplitude and the constraint-derived amplitude inferred from \eqref{eq:aptbm_power}.

Combining the above two operations, the $n$-th coarsely reconstructed block is given by 
\begin{equation}
	\check{\mathbf{c}}_n =   \breve{\mathbf{c}}_n \odot e^{\jmath  \boldsymbol{\psi}_n }, \ \forall n. \label{eq:coarse_reconstruction}
\end{equation}
The resulting  blocks $\left\{\check{\mathbf{c}}_n\right\}$ provide a coarse approximation of the original APTBM  blocks, effectively mitigating the dominant component of nonlinear distortion. Nevertheless, because the amplitude and phase compensations rely on heuristic metrics and are not exact, a residual distortion component remains. Therefore, the coarse reconstruction output can be mathematically formulated as
\begin{equation}
\check{\mathbf{c}}_n = {\mathbf{c}}_n + \mathbf{r}_n, \ \forall n. \label{eq:coarse_reconstruction_error}
\end{equation}
The following subsection presents the fine reconstruction stage, which further refines the coarse reconstruction results.

\subsection{Fine Reconstruction Stage}
\label{sec:LS_SR}

The fine reconstruction stage aims to suppress the residual distortion component $\mathbf{r}_n$ in \eqref{eq:coarse_reconstruction_error} through mathematical optimization. 
Specifically, we first formulate the corresponding nonconvex optimization problem, and subsequently derive a closed-form solution for its reduced-complexity counterpart.

\subsubsection{Problem Formulation}

The APTBM block is governed by two intrinsic constraints: the amplitude constraint in \eqref{eq:aptbm_power} and the phase constraint in \eqref{eq:aptbm_phase}. Although the coarse reconstruction stage explicitly leverages the amplitude constraint, the phase constraint remains underutilized. To further refine the coarse estimates, both constraints must be enforced jointly on a per-block basis to directly minimize the residual distortion, thereby facilitating more accurate signal recovery. Accordingly, the fine reconstruction task for the $n$-th APTBM block is formulated as
\begin{subequations}
	\label{eq:original_SR}
	\begin{align}
		\underset{ {\mathbf{c}}_n }{\text{minimize}} & \quad \left\| \mathbf{r}_n \right\|^2 =  \left\|\check{\mathbf{c}}_{n}-\mathbf{c}_{n}\right\|^{2} \\
		\text {subject to} & \quad \left|a_{n}\right|^{2}+\left|b_{n}\right|^{2}=P, \\
		& \quad \angle \left( a_{n} \right) +\angle \left(b_{n}\right) =2 \phi_{n}. 
	\end{align} 
\end{subequations}%
In this formulation,  while the amplitude constraint is relatively straightforward to handle, the phase constraint is highly nontrivial and mathematically challenging to enforce directly. To address this, we first estimate the initial phase $\phi_n$ for each block, thereby transforming the original optimization problem into a more mathematically tractable form.

\subsubsection{Initial Phase Estimation} 
The accurate estimation of $\phi_n$ is critical, as it fundamentally dictates the formulation and subsequent resolution of the optimization problem. A conventional approach might approximate $\phi_n$ by arithmetically averaging the phases of the received $n$-th block, obtained via the $\arg(\cdot)$ function. However, this method is highly susceptible to phase ambiguity due to the inherent discontinuity of the $\arg(\cdot)$ function, which wraps phases into the principal interval $(-\pi,  \pi]$. If the true phase of $\check{a}_n$ or $\check{b}_n$ falls outside this principal range, the wrapped values can severely skew the arithmetic mean, thereby introducing substantial estimation bias.

\begin{figure}[t]
	\centering
	\includegraphics[width=0.34\textwidth]{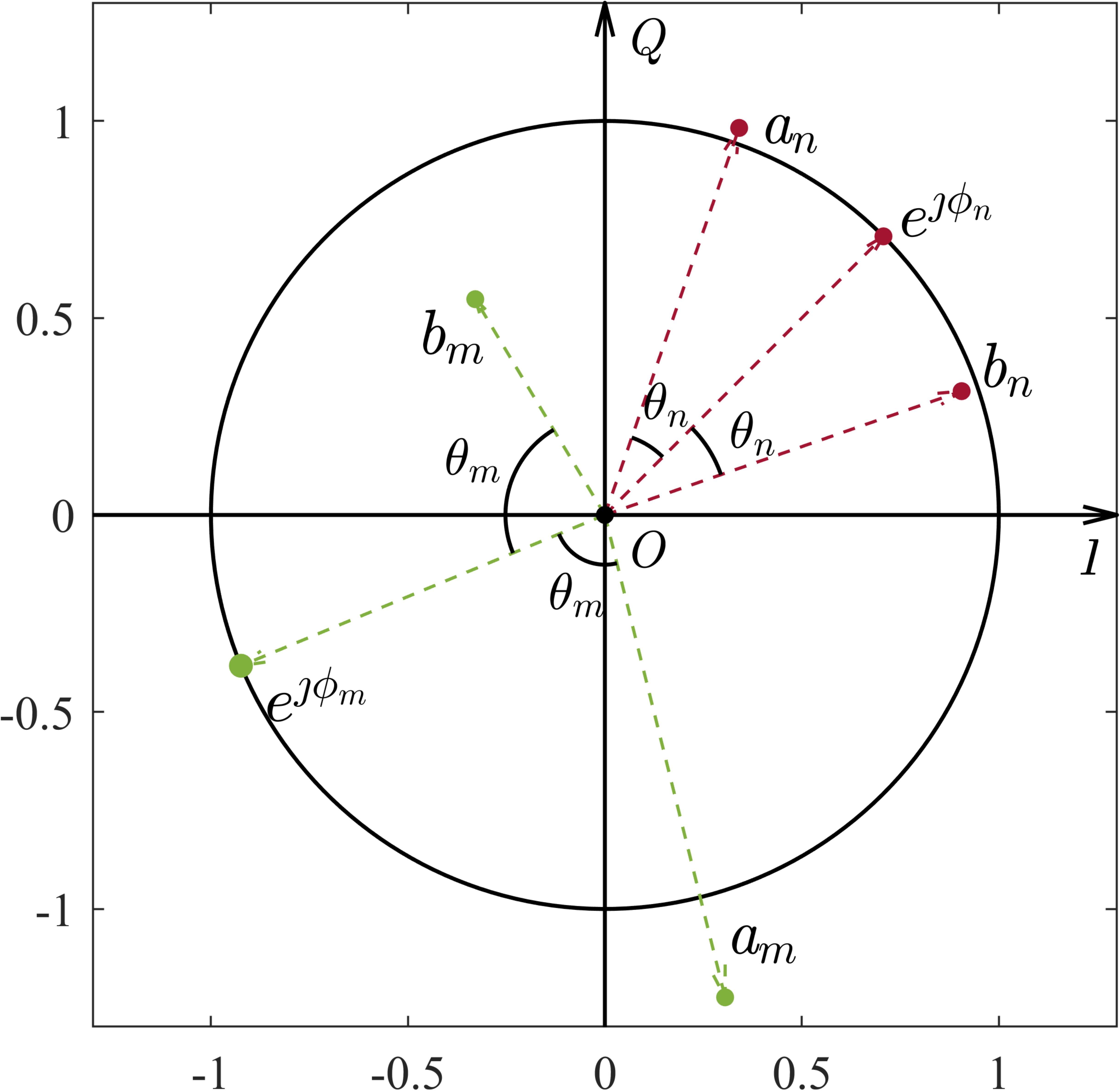}
	\caption{Geometric symmetry of the APTBM blocks.}
	\label{aptbm2}
\end{figure}

To circumvent this limitation, we exploit the inherent geometric symmetry of the APTBM blocks.  As illustrated in Fig.~\ref{aptbm2}, let us consider the $n$-th block  $\mathbf{c}_n$  in the complex plane. On a unit circle centered at the origin $O$, the vectors extending from $O$ to the complex symbols $a_n$ and $b_n$ subtend equal angles  $\theta_n$  with respect to the reference vector $e^{\jmath \phi_n}$. This property dictates that the two symbols within an APTBM block are symmetrically distributed around their associated initial phase, which effectively bisects the angle between them. This symmetry provides an intuitive geometric interpretation of the phase constraint in \eqref{eq:aptbm_phase} and applies universally to all APTBM blocks, as exemplified by the $m$-th block $\mathbf{c}_m$ in Fig.~\ref{aptbm2}. Consequently, rather than arithmetically averaging the potentially wrapped phases, we vectorially combine the unit-normalized symbols in the complex plane and extract the phase of the resultant vector. Formally, the initial phase is estimated as
\begin{equation}
\hat{\phi}_n = \mathcal{G} \left(\arg\left(\frac{\check{a}_n}{\left|\check{a}_n\right|} +
\frac{\check{b}_n}{\left|\check{b}_n\right|}\right)\right),
\label{eq:initial_phase_2}
\end{equation}
where $\mathcal{G}(\cdot)$ denotes a quantization function that maps the resulting  phase to the nearest valid point in the set of possible initial phases. This geometric averaging approach entirely avoids arithmetic operations in the phase domain after the $\arg(\cdot)$ function, thereby eliminating potential phase-wrapping ambiguities. The subsequent quantization step further improves robustness by aligning the estimate with the discrete set of valid initial phases.

Replacing $\phi_n$ with its robust estimate $\hat{\phi}_n$ renders the optimization problem in \eqref{eq:original_SR} tractable. In the following, we leverage this reduced formulation to derive an exact closed-form solution.

\subsubsection{Optimization Solution}
\label{sec:optimization_algorithm}
{Formally, let $a_n = r_{n,1} e^{\jmath \vartheta_{n,1}}$ and $b_n = r_{n,2} e^{\jmath \vartheta_{n,2}}$, where $r_{n,1}, r_{n,2} \geq 0$. By incorporating the estimated initial phase $\hat{\phi}_n$, the optimization problem in \eqref{eq:original_SR} can be reformulated as
\begin{subequations}
\label{problem:23}
\begin{align}
\underset{ \substack{r_{n,1},\ r_{n,2}, \\ \vartheta_{n,1}, \ \vartheta_{n,2}}}{\operatorname{minimize}} \quad &  | \check{a}_n - r_{n,1} e^{\jmath \vartheta_{n,1}} |^2 + | \check{b}_n - r_{n,2} e^{\jmath \vartheta_{n,2}} |^2 \\
\text {subject to} \quad & r_{n,1}^2+r_{n,2}^2=P, \\
& \vartheta_{n,1}+\vartheta_{n,2}=2 \hat{\phi}_n.
\end{align}
\end{subequations}
Substituting the amplitude and phase constraints, namely $r_{n,2} = \sqrt{P - r_{n,1}^2}$ and $\vartheta_{n,2} = 2 \hat{\phi}_n - \vartheta_{n,1}$, into the objective function yields the algebraic expansion shown in \eqref{eq:objective_expansion}.
\begin{figure*}
\begin{equation}
\label{eq:objective_expansion}
\begin{aligned}
& | \check{a}_n - r_{n,1} e^{\jmath \vartheta_{n,1}} |^2 + | \check{b}_n - \sqrt{P - r_{n,1}^2} e^{\jmath (2 \hat{\phi}_n - \vartheta_{n,1})} |^2 \\
& =  r_{n,1}^2 + |\check{a}_n|^2 - r_{n,1} \left( \check{a}_n e^{-\jmath \vartheta_{n,1}} + \check{a}_n^* e^{\jmath \vartheta_{n,1}} \right) + P - r_{n,1}^2 + |\check{b}_n|^2 - \sqrt{P - r_{n,1}^2} \left( \check{b}_n e^{-\jmath (2 \hat{\phi}_n - \vartheta_{n,1})} + \check{b}_n^* e^{\jmath (2 \hat{\phi}_n - \vartheta_{n,1})} \right), \\
& =  P + |\check{a}_n|^2 + |\check{b}_n|^2 -  \left(r_{n,1} \check{a}^*_n + \sqrt{P - r_{n,1}^2} \check{b}_n e^{-\jmath 2 \hat{\phi}_n} \right) e^{\jmath \vartheta_{n,1}} - \left(r_{n,1} \check{a}_n + \sqrt{P - r_{n,1}^2} \check{b}^*_n e^{\jmath 2 \hat{\phi}_n} \right) e^{-\jmath \vartheta_{n,1}}.
\end{aligned}
\end{equation}
\end{figure*}
Since the first three terms in \eqref{eq:objective_expansion} are constant with respect to the optimization variables, minimizing this overall objective is mathematically equivalent to maximizing the sum of the last two terms, as formulated in \eqref{eq:objective_maximization}.
\begin{figure*}
\begin{equation}
\label{eq:objective_maximization}
\underset{  r_{n,1}, \ \vartheta_{n,1}}{\operatorname{maximize}} \quad \left(r_{n,1} \check{a}^*_n + \sqrt{P - r_{n,1}^2} \check{b}_n e^{-\jmath 2 \hat{\phi}_n} \right) e^{\jmath \vartheta_{n,1}} + \left(r_{n,1} \check{a}_n + \sqrt{P - r_{n,1}^2} \check{b}^*_n e^{\jmath 2 \hat{\phi}_n} \right) e^{-\jmath \vartheta_{n,1}}.
\end{equation}
\end{figure*}
To achieve this maximum, the phase $\vartheta_{n,1}$ must perfectly align with the angle of its corresponding complex coefficient. Therefore, the optimal phase $\vartheta_{n,1}^\star$ is straightforwardly given by
\begin{equation}
\vartheta_{n,1}^\star = \angle \left( r_{n,1} \check{a}_n + \sqrt{P - r_{n,1}^2} \check{b}^*_n e^{\jmath 2 \hat{\phi}_n} \right).
\end{equation}
Substituting $\vartheta_{n,1}^\star$ back into the objective function reduces the optimization to a single-variable maximization problem with respect to the amplitude $r_{n,1}$,  expressed as
\begin{equation}
\label{problem:27}
\underset{  r_{n,1} }{\operatorname{maximize}} \quad \left| r_{n,1} \check{a}_n + \sqrt{P - r_{n,1}^2} \check{b}^*_n e^{\jmath 2 \hat{\phi}_n} \right|.
\end{equation}}

{To facilitate a closed-form solution, we introduce a trigonometric substitution by letting $A_n = \check{a}_n$, $B_n = \check{b}^*_n e^{\jmath 2 \hat{\phi}_n}$, $r_{n,1} = \sqrt{P} \cos \alpha_n$, and $\sqrt{P - r_{n,1}^2} = \sqrt{P} \sin \alpha_n$, where the angular variable $\alpha_n \in [0,  \frac{\pi}{2}]$. Consequently, the maximization problem in \eqref{problem:27} can be   reformulated as
\begin{equation}
\label{problem:28}
\underset{\alpha_n \in [0,   \frac{\pi}{2}]}{\operatorname{maximize}} \quad  \left| A_n \cos \alpha_n + B_n \sin \alpha_n \right|^2.
\end{equation}
Expanding the objective function yields the expression shown in \eqref{eq:objective_expansion2}, where $C_n = \cos \left( \angle (A_n) - \angle (B_n) \right)$. By invoking the double-angle formulas, this expression is further streamlined into \eqref{eq:objective_expansion3}.
\begin{figure*}
\begin{subequations}
\begin{align}
\left| A_n \cos \alpha_n + B_n \sin \alpha_n \right|^2 & =  \left|A_n\right|^2  \cos^2 \alpha_n + \left|B_n\right|^2  \sin^2 \alpha_n + 2 \left|A_n\right| \left|B_n\right| C_n \cos \alpha_n \sin \alpha_n, \label{eq:objective_expansion2} \\
& = \frac{\left|A_n\right|^2 + \left|B_n\right|^2}{2} + \frac{\left|A_n\right|^2 - \left|B_n\right|^2}{2} \cos 2 \alpha_n + \left|A_n\right| \left|B_n\right| C_n \sin 2 \alpha_n. \label{eq:objective_expansion3}
\end{align}
\end{subequations}
\end{figure*}}

{To decouple the trigonometric terms, we introduce a new set of variables: $\beta_n = 2 \alpha_n \in [0,  \pi]$, $D_n = \frac{\left|A_n\right|^2 + \left|B_n\right|^2}{2}$, $E_n = \frac{\left|A_n\right|^2 - \left|B_n\right|^2}{2}$, and $F_n = \left|A_n\right| \left|B_n\right| C_n$. This allows us to recast the problem as
\begin{equation}
\label{problem:30}
\underset{\beta_n \in [0, \pi]}{\operatorname{maximize}} \quad   D_n +  E_n \cos \beta_n + F_n \sin \beta_n.
\end{equation}
Finally, applying the harmonic addition theorem, we  merge the sine and cosine components into a single phase-shifted cosine function, as expressed in  \eqref{eq:objective_expansion4}, where $\cos \gamma_n = \frac{E_n}{\sqrt{E_n^2 + F_n^2}}$ and $\sin \gamma_n = \frac{F_n}{\sqrt{E_n^2 + F_n^2}}$.
\begin{figure*}
\begin{equation}
\label{eq:objective_expansion4}
\begin{aligned}
D_n + E_n \cos \beta_n + F_n \sin \beta_n & = D_n + \sqrt{E_n^2 + F_n^2} \left( \frac{E_n}{\sqrt{E_n^2 + F_n^2}} \cos \beta_n + \frac{F_n}{\sqrt{E_n^2 + F_n^2}} \sin \beta_n \right), \\
& = D_n + \sqrt{E_n^2 + F_n^2} \cos \left( \beta_n - \gamma_n \right).
\end{aligned}
\end{equation}
\end{figure*}}

{To maximize the objective function in \eqref{problem:30}, we analyze the behavior of the phase shift $\gamma_n$ over the feasible domain $\beta_n \in [0, \pi]$.  This leads to two distinct cases  based on the sign of   $C_n$:
\begin{itemize}
\item If $C_n \geq 0$, i.e., $\sin \gamma_n \geq 0$, the unconstrained peak $\gamma_n$ falls within the feasible region $[0,  \pi]$. Consequently, the optimal $\beta_n$ that maximizes the objective is an interior point, given by
\begin{equation}
\beta_n^\star = \gamma_n = \mathrm{atan2}\Big(F_n, E_n\Big).
\end{equation}
\item If $C_n < 0$, i.e., $\sin \gamma_n < 0$, the unconstrained peak lies outside the feasible region $[0, \pi]$. Given the monotonic behavior of the cosine function  within this restricted range, the objective function is maximized at one of the boundaries, i.e., $\beta_n \in \{0, \pi\}$. 
Specifically, if $\beta_n^\star = 0$, we have  
\begin{equation}
D_n + E_n \cos \beta_n^\star + F_n \sin \beta_n^\star = \left|A_n\right|^2. 
\end{equation} 
If $\beta_n^\star = \pi$, we have 
\begin{equation}
D_n + E_n \cos \beta_n^\star + F_n \sin \beta_n^\star = \left|B_n\right|^2.
\end{equation}
By comparing these boundary values, the optimal $\beta_n$ is determined as
\begin{equation}
\beta_n^\star = \left\{
\begin{array}{cl}
0, & \text{if } \left|A_n\right|^2 \geq \left|B_n\right|^2, \\
\pi, & \text{otherwise}.  
\end{array}
\right.
\end{equation}
\end{itemize}}

{By synthesizing these analytical findings and mapping them back to the original variables, we can explicitly express the optimal parameters $r_{n,1}^\star$, $r_{n,2}^\star$, $\vartheta_{n,1}^\star$, and $\vartheta_{n,2}^\star$ as
\begin{equation}
\label{problem:36}
\begin{aligned}
r_{n,1}^\star &= \sqrt{P} \cos \alpha_n^\star, \\
r_{n,2}^\star &= \sqrt{P} \sin \alpha_n^\star, \\
\vartheta_{n,1}^\star &= \angle \left( r_{n,1}^\star \check{a}_n + r_{n,2}^\star \check{b}^*_n e^{\jmath 2 \hat{\phi}_n} \right), \\
\vartheta_{n,2}^\star &= 2 \hat{\phi}_n - \vartheta_{n,1}^\star,
\end{aligned}
\end{equation}
where the   angular variable $\alpha_n^\star$ is given by
\begin{equation}
\alpha_n^\star = \left\{
\begin{array}{cl}
\frac{1}{2} \mathrm{atan2}\Big(F_n, E_n\Big), & \text{if } C_n \geq 0, \\
0, & \text{else if } \left|A_n\right|^2 \geq \left|B_n\right|^2, \\
\frac{\pi}{2}, & \text{otherwise}.
\end{array}
\right.
\end{equation}
Finally, the   reconstructed APTBM block is assembled as
\begin{equation}
\label{problem:38}
\hat{\mathbf{c}}_n = \left[ r_{n,1}^\star e^{\jmath \vartheta_{n,1}^\star},  r_{n,2}^\star e^{\jmath \vartheta_{n,2}^\star} \right]^T.
\end{equation}}

\begin{algorithm}[t]
	\caption{Two-Stage Signal Reconstruction Algorithm}
	\label{alg:Proposed_SR}
	\begin{algorithmic}[1]
		\STATE \textbf{Input:} Received APTBM blocks $\left\{\tilde{\mathbf{c}}_n\right\}$,
							   phase shift LUT $\phi_{\mathrm{a}}(\cdot)$, and initial phase alphabet.
		\STATE \textbf{Output:} Reconstructed APTBM  blocks $\left\{\hat{\mathbf{c}}_n\right\}$.
		\FOR{$n=1:N$}
			\STATE Compute the coarsely reconstructed block $\check{\mathbf{c}}_n$ via \eqref{eq:coarse_reconstruction}.
			\STATE Estimate the initial phase $\hat{\phi}_n$ using \eqref{eq:initial_phase_2}.
			\STATE Calculate the optimal parameters via the closed-form solution in \eqref{problem:36}, and assemble the finely reconstructed block $\hat{\mathbf{c}}_n$ according to \eqref{problem:38}.
		\ENDFOR
		\STATE \textbf{Return} $\left\{\hat{\mathbf{c}}_n\right\}$.
	\end{algorithmic}
\end{algorithm}

\subsection{Algorithm Overview}
\label{sec:proposed_SR_overview}
The overall procedure of the proposed two-stage signal reconstruction algorithm is summarized in Algorithm~\ref{alg:Proposed_SR}. 
The computational complexities for the coarse reconstruction, initial phase estimation, and   optimization solution per APTBM  block are $\mathcal{O}(1)$,  $\mathcal{O}(M)$, and $\mathcal{O}(1)$, respectively. 
Consequently, the overall computational complexity of the proposed algorithm is  $\mathcal{O}(NM)$.

Even though the initial phase estimation introduces additional computational overhead, it provides a crucial advantage by drastically reducing the complexity of the subsequent demodulation step.  Specifically, because the initial phases are inherently resolved during the reconstruction stage, the demodulator can directly map the reconstructed symbols to the nearest constellation points on the spherical surface without executing an exhaustive search over all candidate phases. This reduces the demodulation complexity from $\mathcal{O}(ML)$ to $\mathcal{O}(L)$ per block. As a result, the aggregate computational complexity of the proposed framework remains exceptionally low, ensuring its viability for real-world deployment.

\section{Performance Evaluation}
\label{sec:simulation}
In this section, we evaluate the performance of the proposed signal reconstruction 
algorithm using both numerical simulations and testbed experiments.

\subsection{Numerical Simulations}
\label{sec:simulation_num}
In the simulations, the PA nonlinear distortion is modeled using the modified Rapp model defined in \eqref{eq:modified_rapp}, with the fitting parameters listed in Table~\ref{tab:PA_parameters} \cite{fanAmplitudephasetimeBlockModulation2023}.
The  input saturation power of this model is approximately -5 dBm. The   channel is  assumed to be an AWGN channel, and the signal-to-noise ratio (SNR) is fixed at 30 dB, unless otherwise specified.
A root-raised-cosine (RRC) filter with a roll-off factor of 0.25 and an oversampling factor of 4 is employed for pulse shaping. The signal reconstruction algorithm  in \cite{fanAmplitudephasetimeBlockModulation2023} is adopted as the baseline for comparison, while conventional quadrature amplitude modulation (QAM) is also included as a benchmark.  All simulation results are averaged over $10^6$ Monte Carlo trials.

\begin{table}[t]
	\centering
	\caption{Parameter Settings of the Modified Rapp Model}
	\label{tab:PA_parameters}
	\begin{tabular}{|c|c|c|c|c|c|c|c|}
		\hline
		\textbf{Parameters} & $g_0$ & $A_{\text{sat}}$ & $\alpha_0$ & $\beta_0$ & $q_0$ & $q_1$ & $q_2$ \\ \hline
		\textbf{Values} & 4.65  & 0.58   & 2560   & 0.114   & 0.81 & 2.4 & 2.3    \\ \hline
	\end{tabular}
\end{table}

The bit error rate (BER) performance under varying IBO levels is shown in Fig.~\ref{fig:simulation_power}. 
A smaller IBO improves PAE but also introduces stronger nonlinear distortion, thereby degrading the BER performance.
It can be observed that the proposed algorithm consistently outperforms both the baseline and conventional QAM schemes. At a target BER of $10^{-4}$ with a modulation order (MO) of 64, the baseline achieves an approximate 6 dB reduction in IBO compared to conventional QAM. Notably, the proposed algorithm further reduces the IBO by an additional 2 dB relative to the baseline. {Based on the efficiency formulation in \eqref{eq:PAE} and the PA   characteristics modeled in \eqref{eq:modified_rapp}, this 2 dB reduction in IBO translates to a substantial relative PAE improvement of approximately 39.3\%.}
For a more relaxed BER target of $10^{-3}$, representing more severe nonlinear operating conditions,  the proposed algorithm achieves roughly a 5 dB additional IBO reduction over the baseline, translating to a 77.8\% PAE enhancement. 
{The baseline algorithm suffers a sharp performance degradation when the IBO drops below a critical threshold, whereas the proposed method exhibits remarkable stability. This divergent behavior is primarily due to the baseline's high sensitivity to phase distortion, which triggers a severe threshold effect characterized by a rapid surge in BER once a specific distortion level is exceeded. In contrast, the proposed algorithm leverages a dedicated phase compensation mechanism to effectively neutralize such distortions, resulting in a much more graceful BER degradation.}

\begin{figure}
	\centering
	\includegraphics[width=0.43\textwidth]{./figures/power_vs_BER.pdf}
	\caption{BER comparisons  with varying IBO.}
	\label{fig:simulation_power}
\end{figure}

To ensure a rigorous and fair comparison, the baseline algorithm is augmented with the identical phase compensation (PC) mechanism employed by our proposed method. Subsequent simulations are conducted to evaluate the true performance impact of this specific enhancement. As depicted in Fig.~\ref{fig:simulation_power2}, at a target BER of $10^{-4}$, the proposed algorithm achieves a further IBO reduction of over 1 dB relative to the PC-enhanced baseline across various MOs.    
Moreover, this performance gain widens to a substantial 3 dB when operating with an MO of 16. These results demonstrate that the proposed algorithm suppresses nonlinear distortion well beyond the capabilities of PC alone, thereby firmly validating its superior effectiveness.

\begin{figure}
	\centering
	\includegraphics[width=0.43\textwidth]{./figures/power_vs_BER2.pdf}
	\caption{BER gains with varying IBO.}
	\label{fig:simulation_power2}
\end{figure}

\begin{figure}[t]
	\centering
	\subfloat[]{\includegraphics[width=0.48\linewidth]{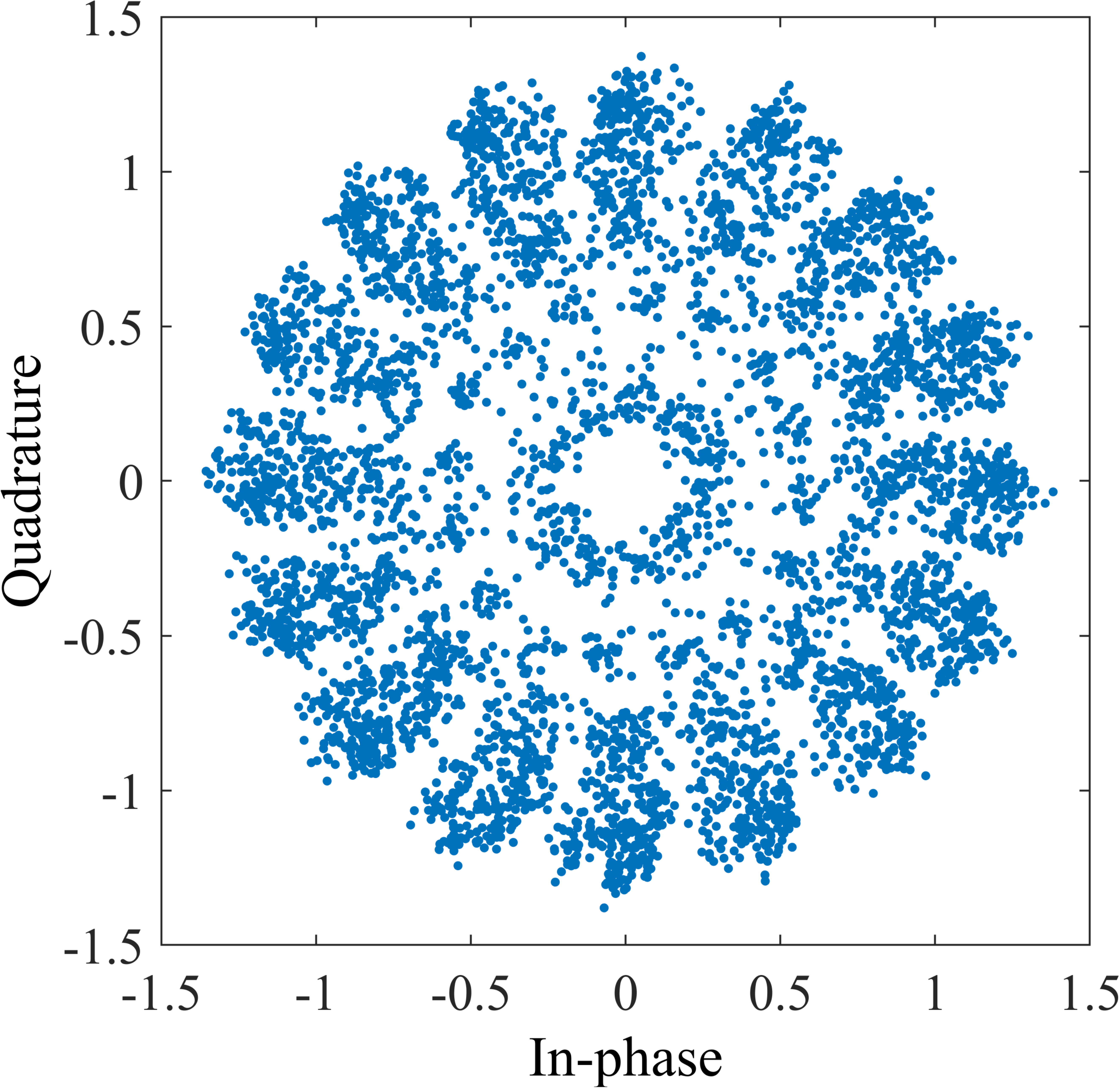} \label{distortion}}%
	\centering
	\subfloat[]{\includegraphics[width=0.48\linewidth]{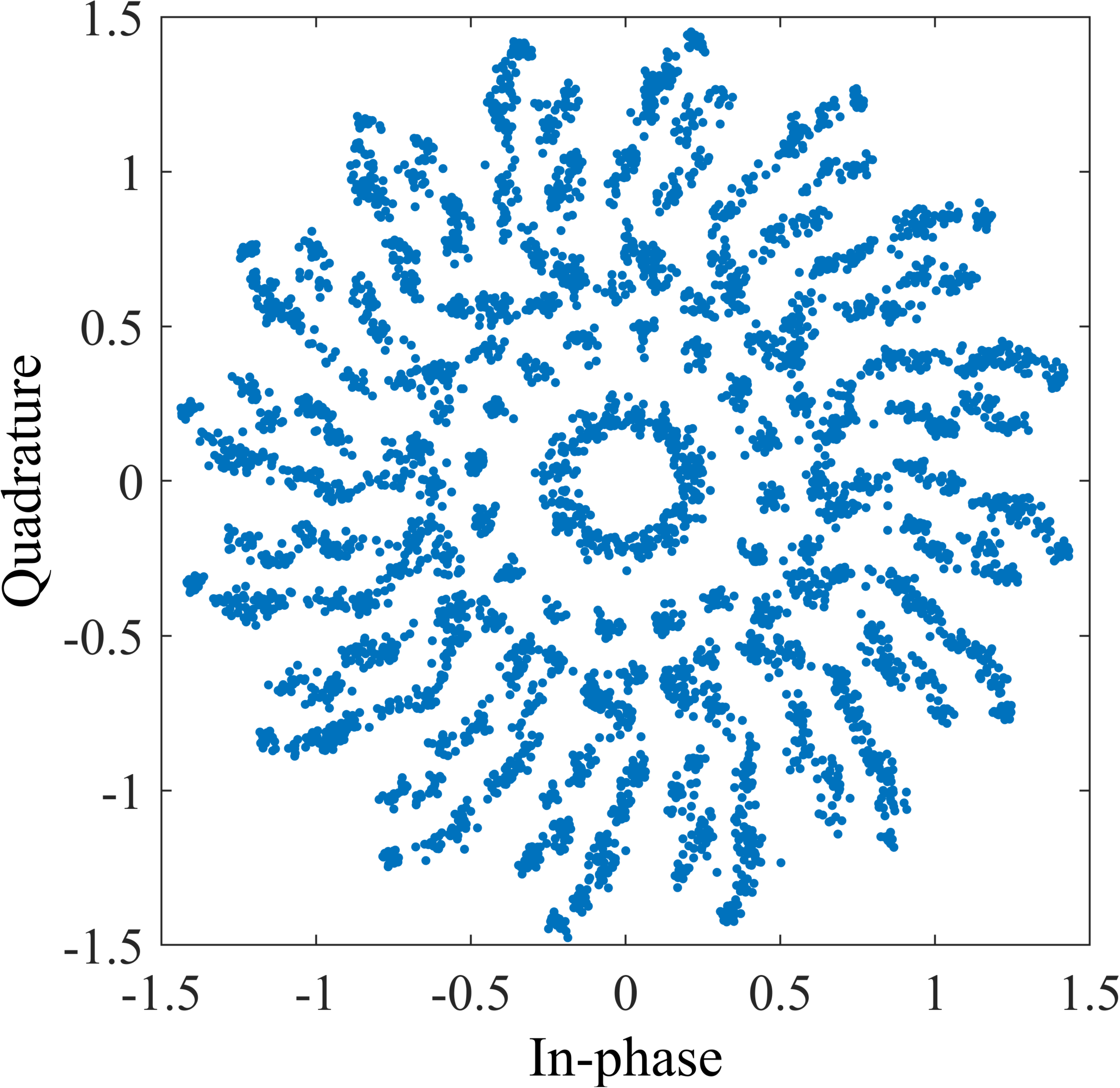} \label{baseline}}%
	
	\centering
	\subfloat[]{\includegraphics[width=0.48\linewidth]{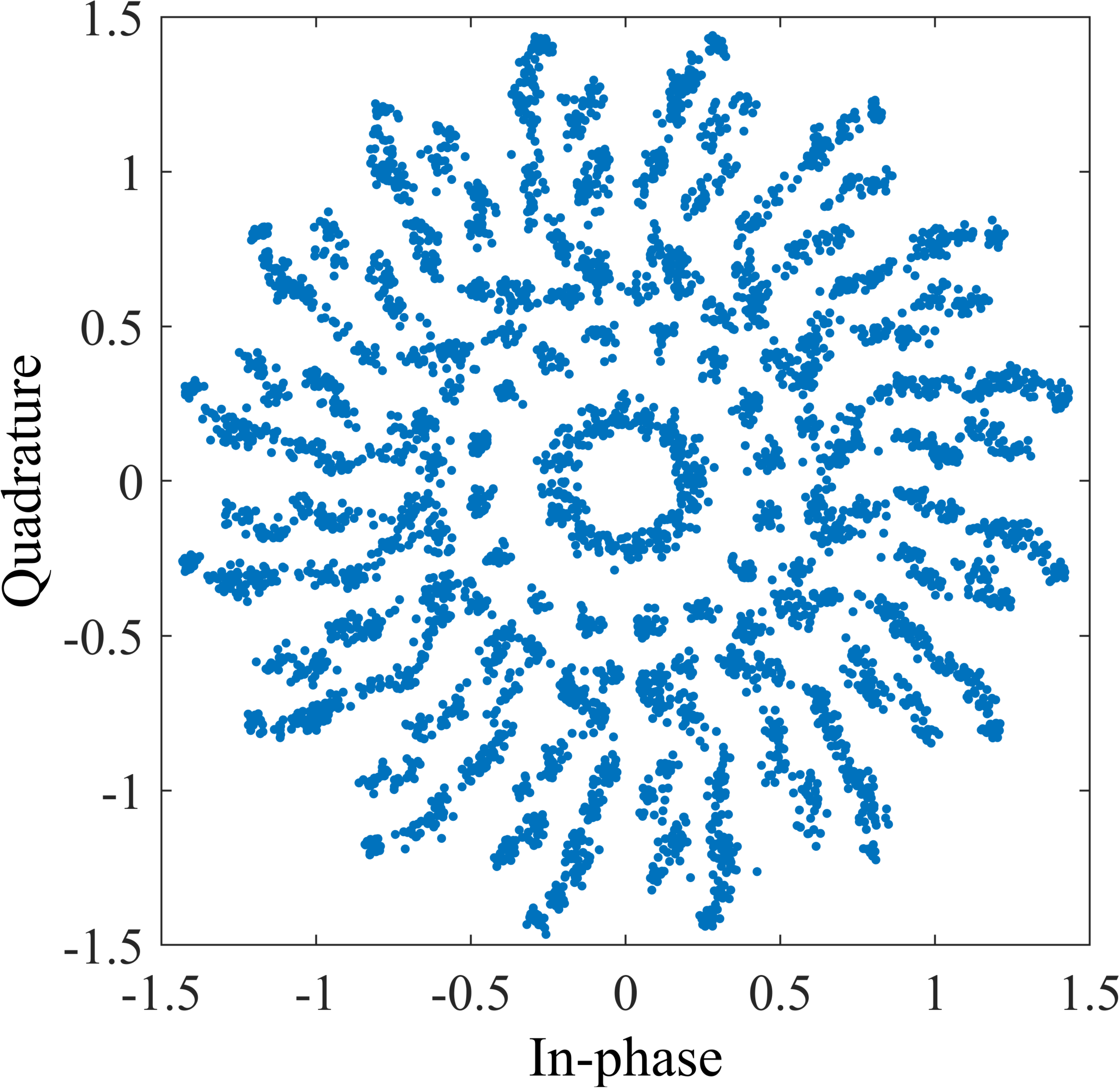} \label{prb}}%
	\centering
	\subfloat[]{\includegraphics[width=0.48\linewidth]{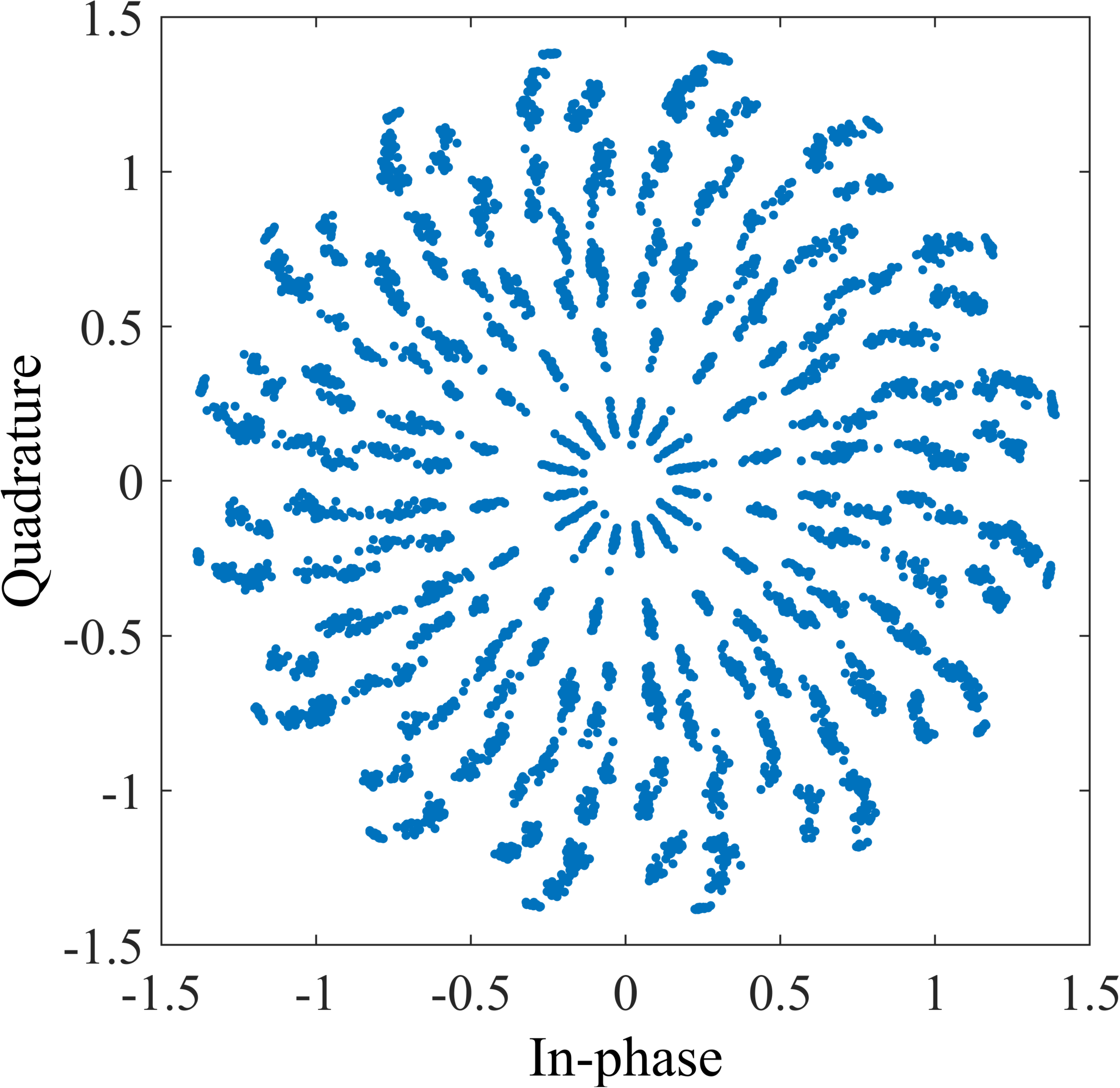} \label{proposed}}%
	\caption{Simulated constellations. (a) Without signal reconstruction. (b) With baseline reconstruction. (c) 
	With PC-enhanced baseline reconstruction. (d) With proposed reconstruction.}
	\label{fig:simulation_constellation}%
\end{figure}

{The simulated constellations  are shown in Fig.~\ref{fig:simulation_constellation}. The PA input power is set to -15 dBm for an MO of 16, corresponding to an IBO of 10 dB.  As observed, the received constellation without signal reconstruction suffers from severe PA-induced nonlinear distortion. While the baseline algorithm provides a noticeable improvement over the unreconstructed case, the proposed algorithm achieves superior performance. Specifically, the proposed method further minimizes the intra-cluster symbol spread, resulting in reconstructed symbols that are much more tightly clustered around their ideal reference points. These results visually corroborate the effectiveness of the proposed algorithm in mitigating nonlinear distortion and enhancing overall signal quality.}

\begin{figure}[t]
	\centering
	\includegraphics[width=0.43\textwidth]{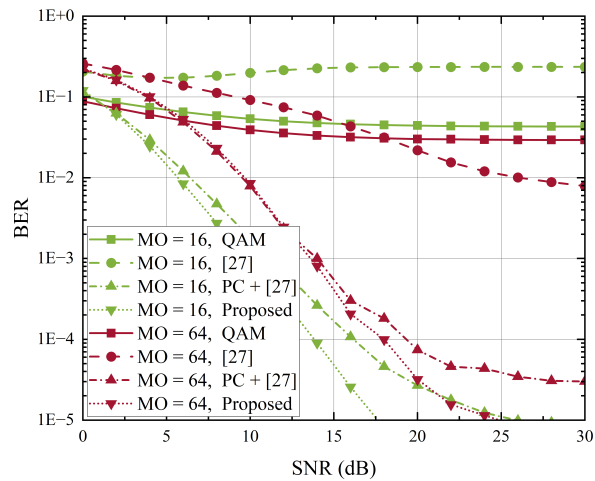}
	\caption{BER comparisons with varying SNR.}
	\label{fig:simulation_nonlinear}
\end{figure}

Beyond reducing the required IBO, the proposed algorithm significantly enhances transmission reliability under fixed IBO and varying SNR conditions. To evaluate the algorithm's robustness against severe nonlinear distortion, the IBO is deliberately set to 8 dB for an MO of 16 and 16 dB for an MO of 64.  As depicted in Fig.~\ref{fig:simulation_nonlinear}, both conventional QAM and the  baseline algorithm suffer from severe performance degradation under these stringent  conditions, with BER levels exceeding $10^{-2}$ even at high SNRs. This behavior occurs because nonlinear distortion dominates the error performance in this regime, and existing methods fail to mitigate such effects effectively.
In contrast, the proposed algorithm and the PC-enhanced baseline   successfully sustain reliable communication performance, reaching BER levels as low as $10^{-5}$.  Notably, the proposed algorithm consistently outperforms the PC-enhanced baseline   across all SNRs, particularly in the high-SNR regime where  nonlinear distortion becomes the primary performance bottleneck.

\subsection{Testbed Experiments}
\label{sec:simulation_hardware}
\begin{figure*}[t]
	\centering
	\includegraphics[width=0.85\textwidth]{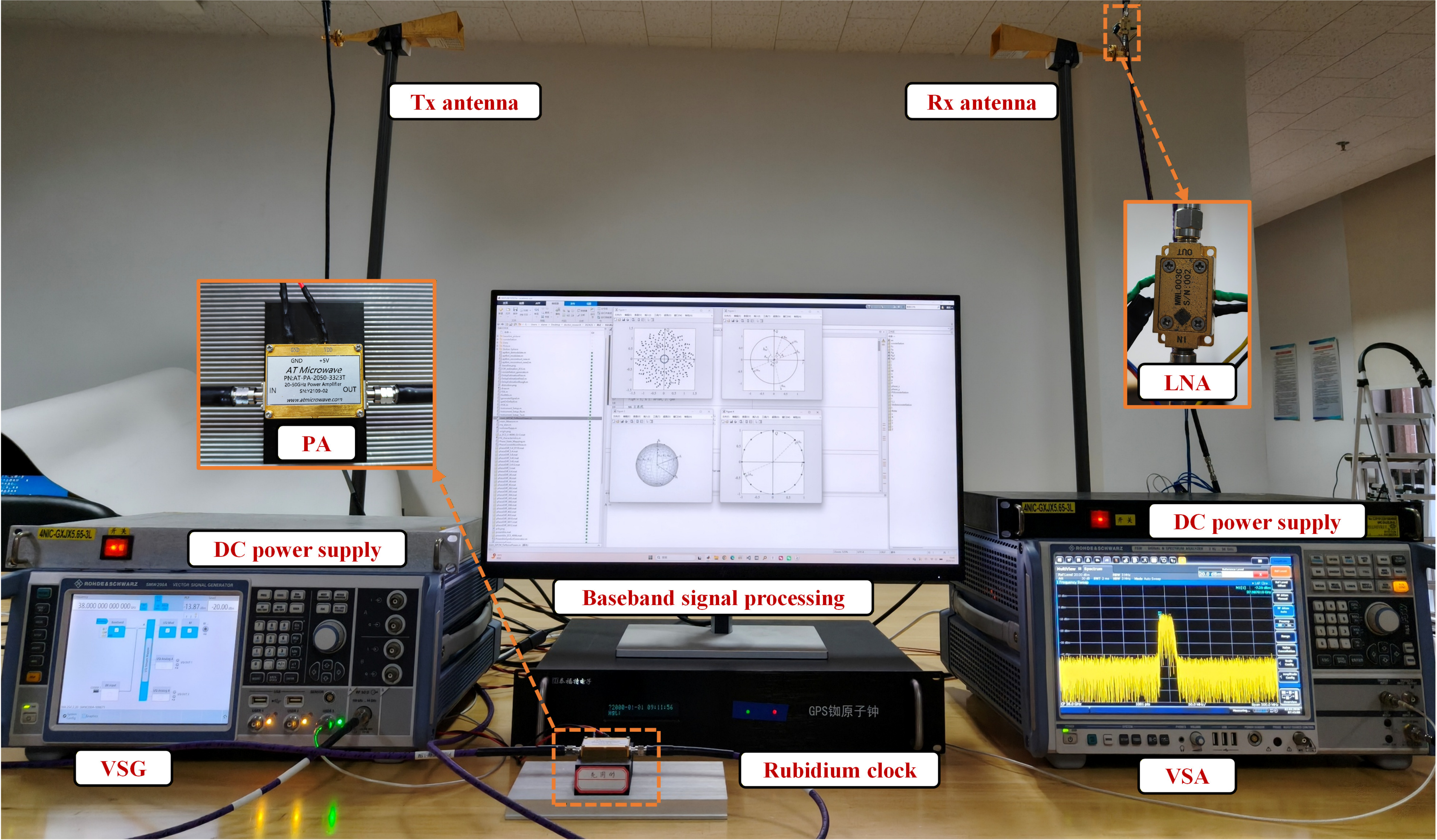}
	\caption{The used experimental platform.}
	\label{fig:hardware_platform}
\end{figure*}

To further validate the proposed algorithm, testbed experiments are conducted using the experimental platform illustrated in Fig.~\ref{fig:hardware_platform}. The  platform consists of a vector signal generator (VSG) {(Rohde \& Schwarz SMW200A)},
a vector signal analyzer (VSA) {(Rohde \& Schwarz FSW50)},
a PA, a low noise amplifier (LNA), Tx and Rx antennas, a rubidium clock, two DC power supplies,  and a central computer. The computer controls the VSG and VSA and performs all baseband signal processing in MATLAB. In the transmit chain, the VSG upconverts the baseband signal to the desired RF carrier and regulates the PA input power. Following over-the-air transmission via the Tx and Rx antennas, the VSA downconverts  the received RF signal back  to baseband. The rubidium clock provides a stable 10 MHz reference signal and a 1 pulse per second (PPS) signal to both the VSG and VSA, ensuring synchronous triggering and carrier synchronization. The LNA amplifies the received signal to enhance the VSA's reception sensitivity, while the DC power supplies provide the necessary voltages to the PA and LNA.

Experiments are conducted in both the sub-6 GHz band (at 5.4 GHz) and the mmWave band (at 38 GHz). Because hardware components are inherently frequency-dependent, different PAs and antennas are employed for each band. {Specifically, for the sub-6 GHz setup, we employ the MWZ038C PA alongside omnidirectional dipole antennas. For the mmWave configuration, we utilize the AT-PA-2050-3323T PA paired with highly directional horn antennas (model HD-400SGAH25+V), whose high directivity effectively compensates for the severe free-space path loss experienced at mmWave frequencies.} All baseband signal processing parameters remain consistent with those used in the numerical simulations. {For channel estimation, a least squares (LS) approach based on known pilot symbols is utilized. To ensure distortion-free channel estimation, the pilot power is strictly restricted to the linear operating region of the PA.} Finally, the symbol error rate (SER) is adopted as the primary performance metric, with the reported results averaged over $10^5$ independent transmitted blocks.

\begin{figure}[t]
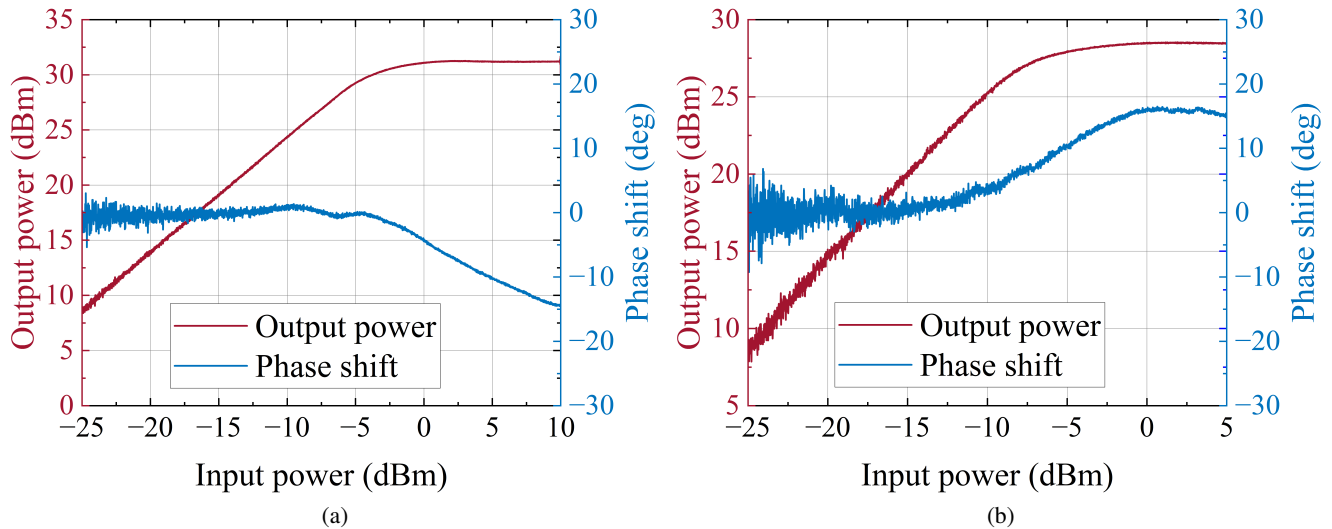

	\centering
	\subfloat[]{\includegraphics[width=0.48\linewidth]{./figures/PA_5.4.pdf} \label{PA_characteristics1}}%
	\centering
	\subfloat[]{\includegraphics[width=0.48\linewidth]{./figures/PA_40.pdf} \label{PA_characteristics2}}%
	\caption{Measured PA characteristics. (a) 5.4 GHz. (b) 38 GHz.}
	\label{fig:hardware_PA_characteristics}
\end{figure}

First, we measure the PA characteristics in both the sub-6 GHz and mmWave bands. The results are depicted in Fig.~\ref{fig:hardware_PA_characteristics}. It can be observed that the PA input saturation power is approximately 1 dBm at 5.4 GHz and -5 dBm at 38 GHz. A closer examination reveals that nonlinear distortion at 5.4 GHz is primarily amplitude-dominated, as the phase shift at saturation  is only about $-5^\circ$. In contrast, nonlinear distortion at 38 GHz exhibits both amplitude compression and significant phase deviation, reaching a phase shift of roughly $15^\circ$ at saturation. These observations indicate that the mmWave band experiences more severe nonlinear distortion than the sub-6 GHz band in our testbed experiments. As established in the previous simulations, the proposed algorithm demonstrates exceptional performance under severe nonlinear conditions. Therefore, the testbed experiments are expected to  highlight its effectiveness in the mmWave band. Finally, the measured AM-PM characteristics are stored in a LUT for use in subsequent experiments. It is worth noting that the AM-PM characteristics of PAs are measured only once during the experiments, as the proposed algorithm does not require highly precise AM-PM characterization.

\begin{figure}[t]
	\centering
	\subfloat[]{\includegraphics[width=0.48\linewidth]{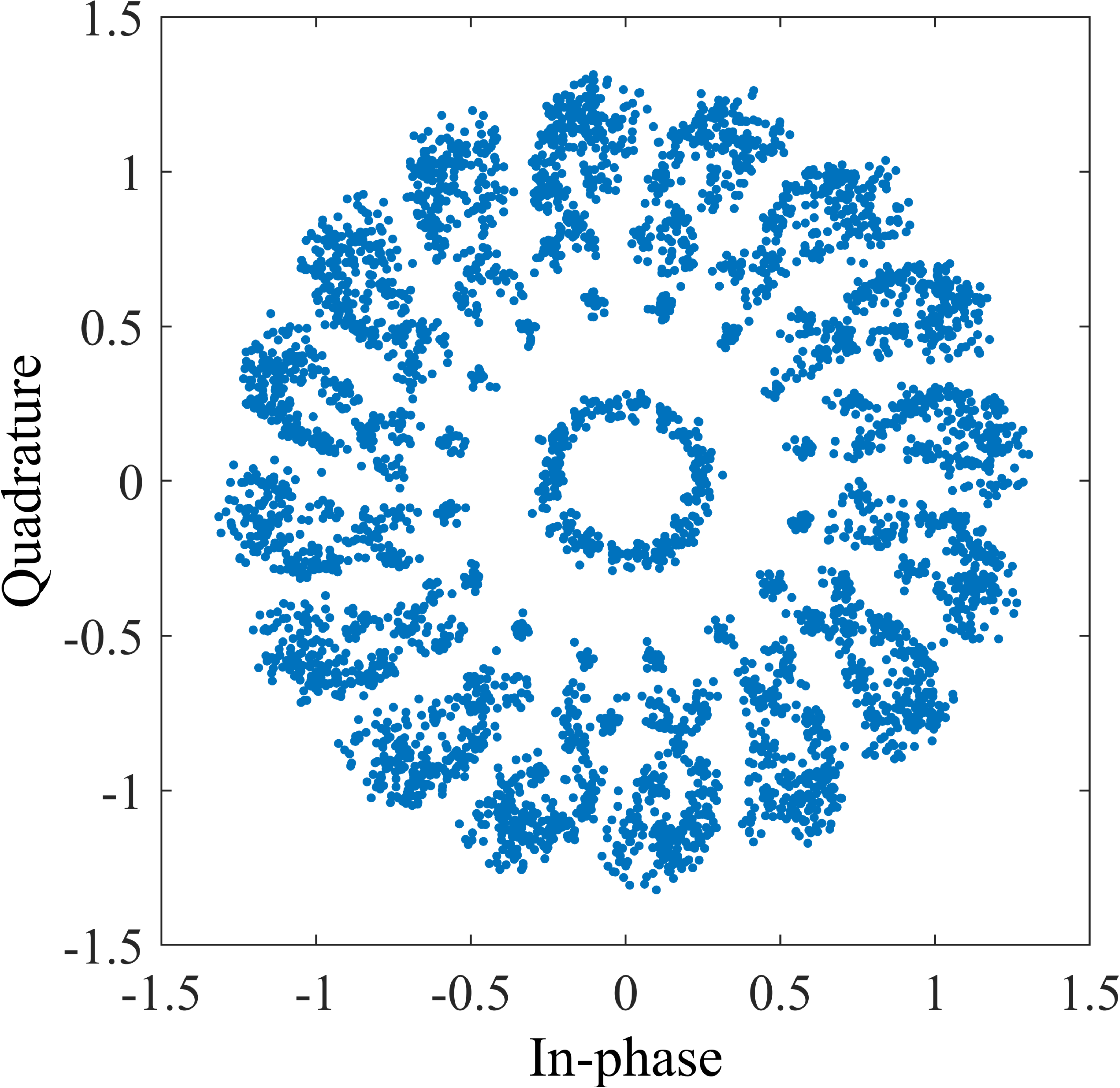} \label{distortion1}}%
	\centering
	\subfloat[]{\includegraphics[width=0.48\linewidth]{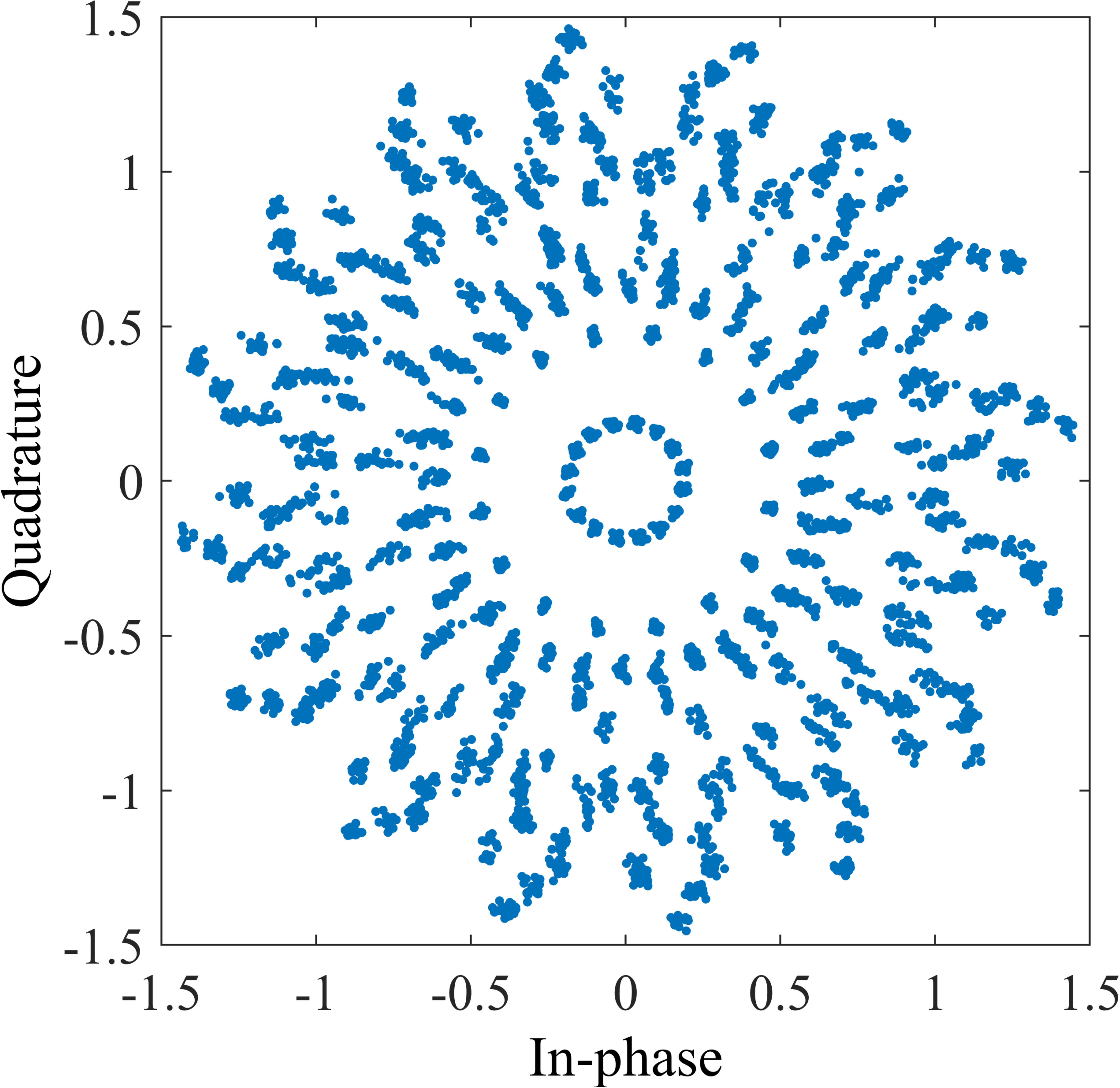} \label{baseline1}}%
	
	\centering
	\subfloat[]{\includegraphics[width=0.48\linewidth]{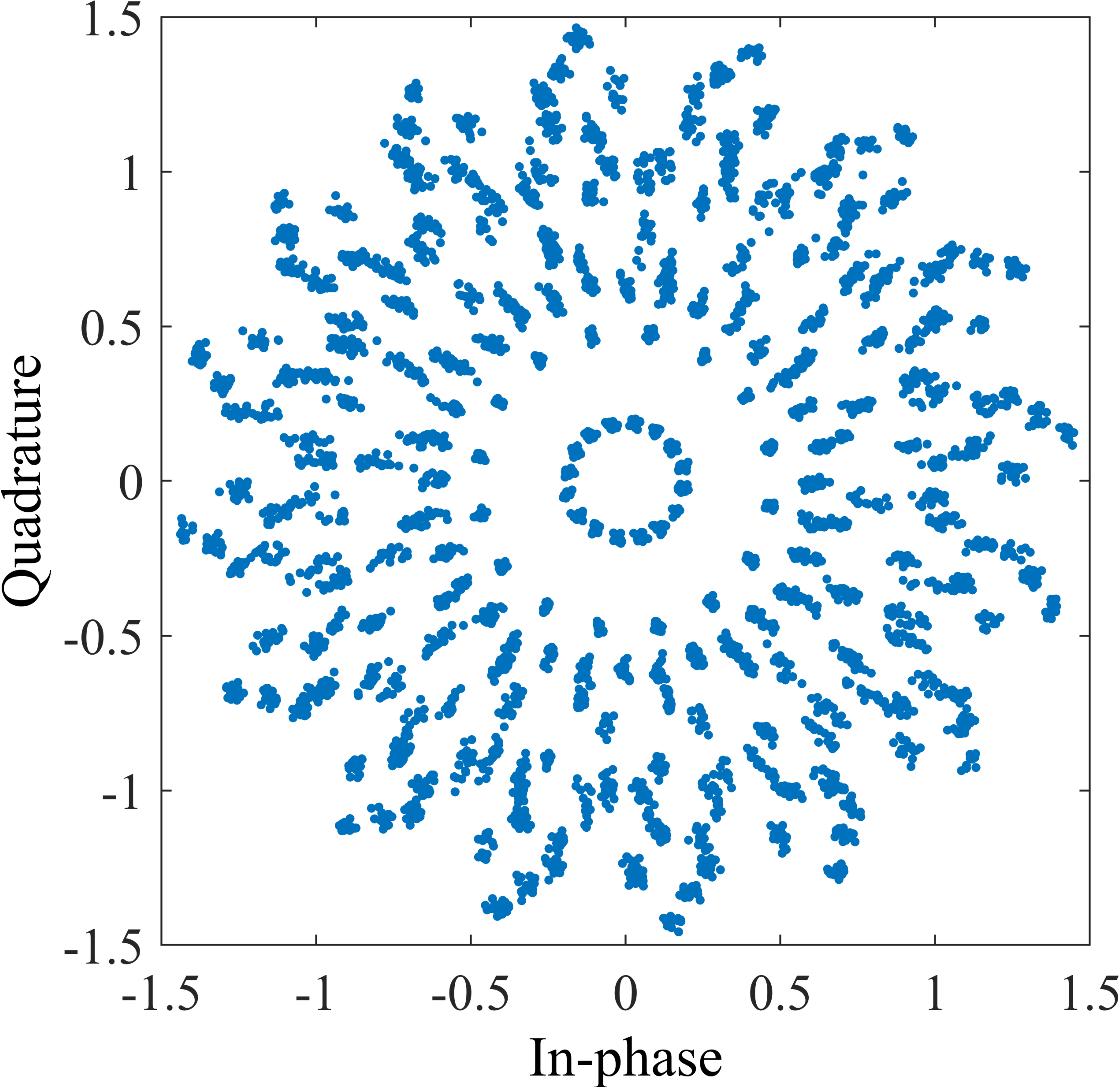} \label{prb1}}%
	\centering
	\subfloat[]{\includegraphics[width=0.48\linewidth]{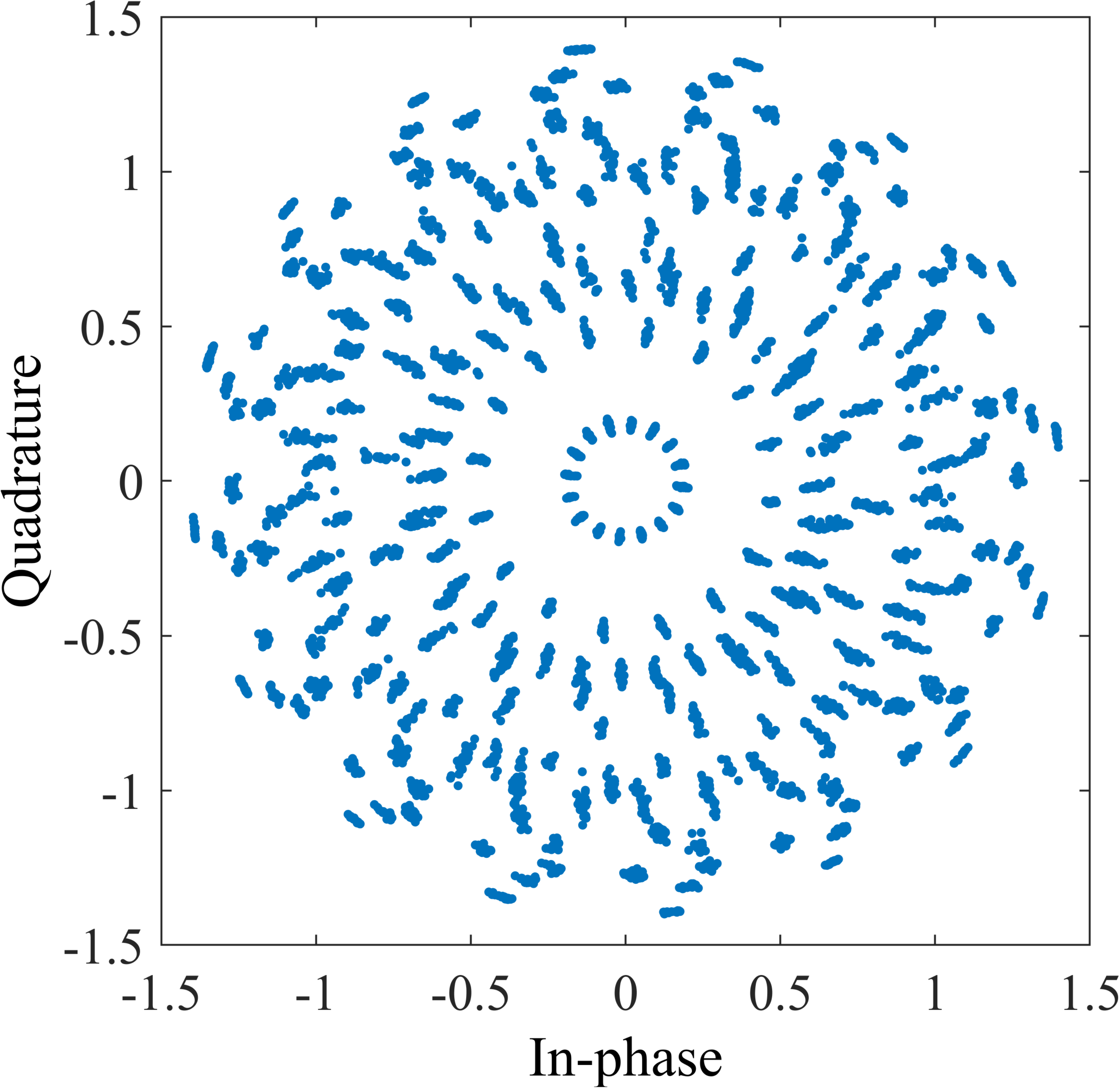} \label{proposed1}}%
	\caption{Measured constellations at 5.4 GHz. (a) Without signal reconstruction. (b) With baseline reconstruction. (c) 
	With PC-enhanced baseline reconstruction. (d) With proposed reconstruction.}
	\label{fig:hardware_constellation1}
\end{figure}

\begin{figure}[t]
	\centering
	\subfloat[]{\includegraphics[width=0.48\linewidth]{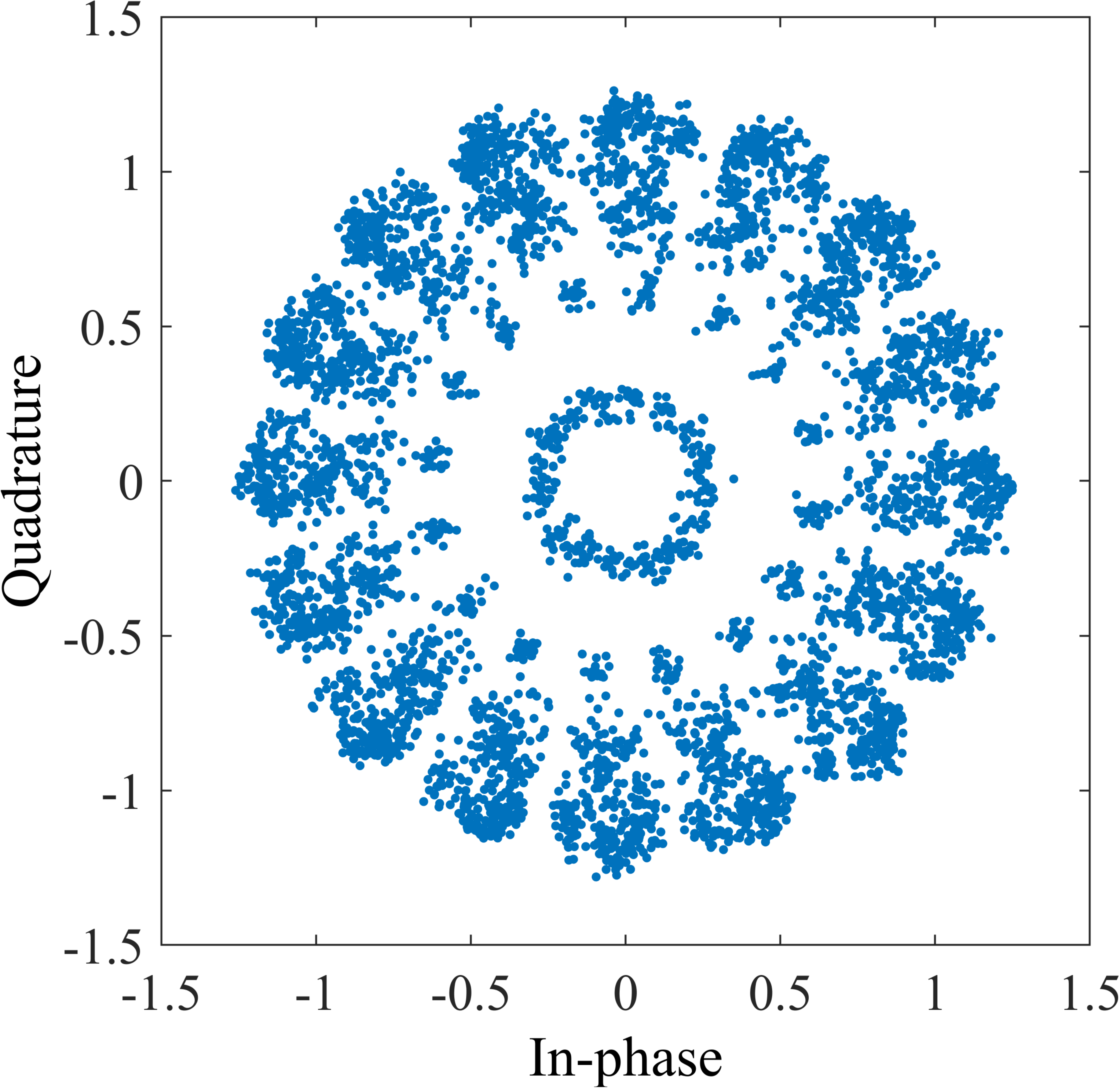} \label{distortion2}}%
	\centering
	\subfloat[]{\includegraphics[width=0.48\linewidth]{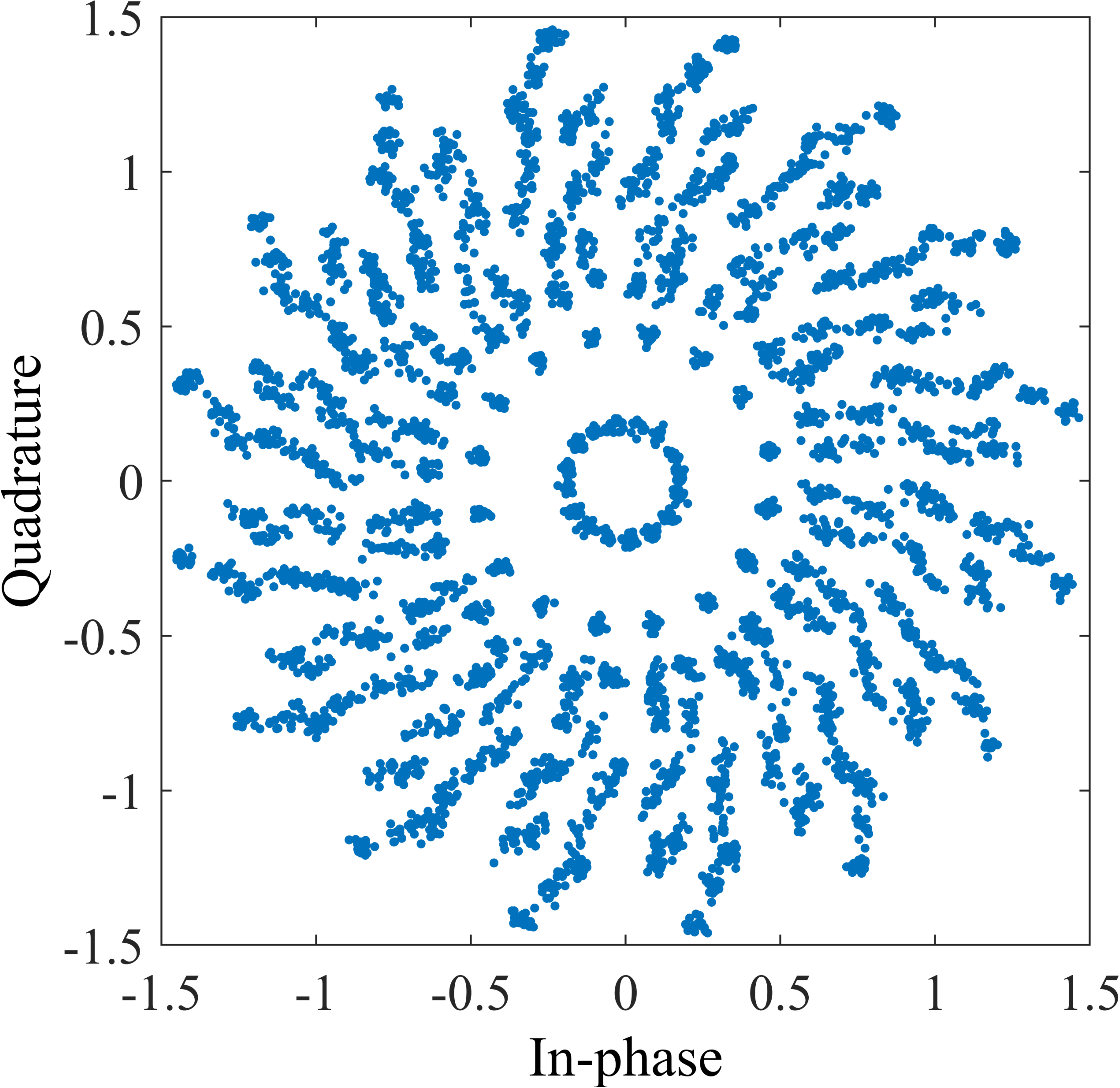} \label{baseline2}}%

	\centering
	\subfloat[]{\includegraphics[width=0.48\linewidth]{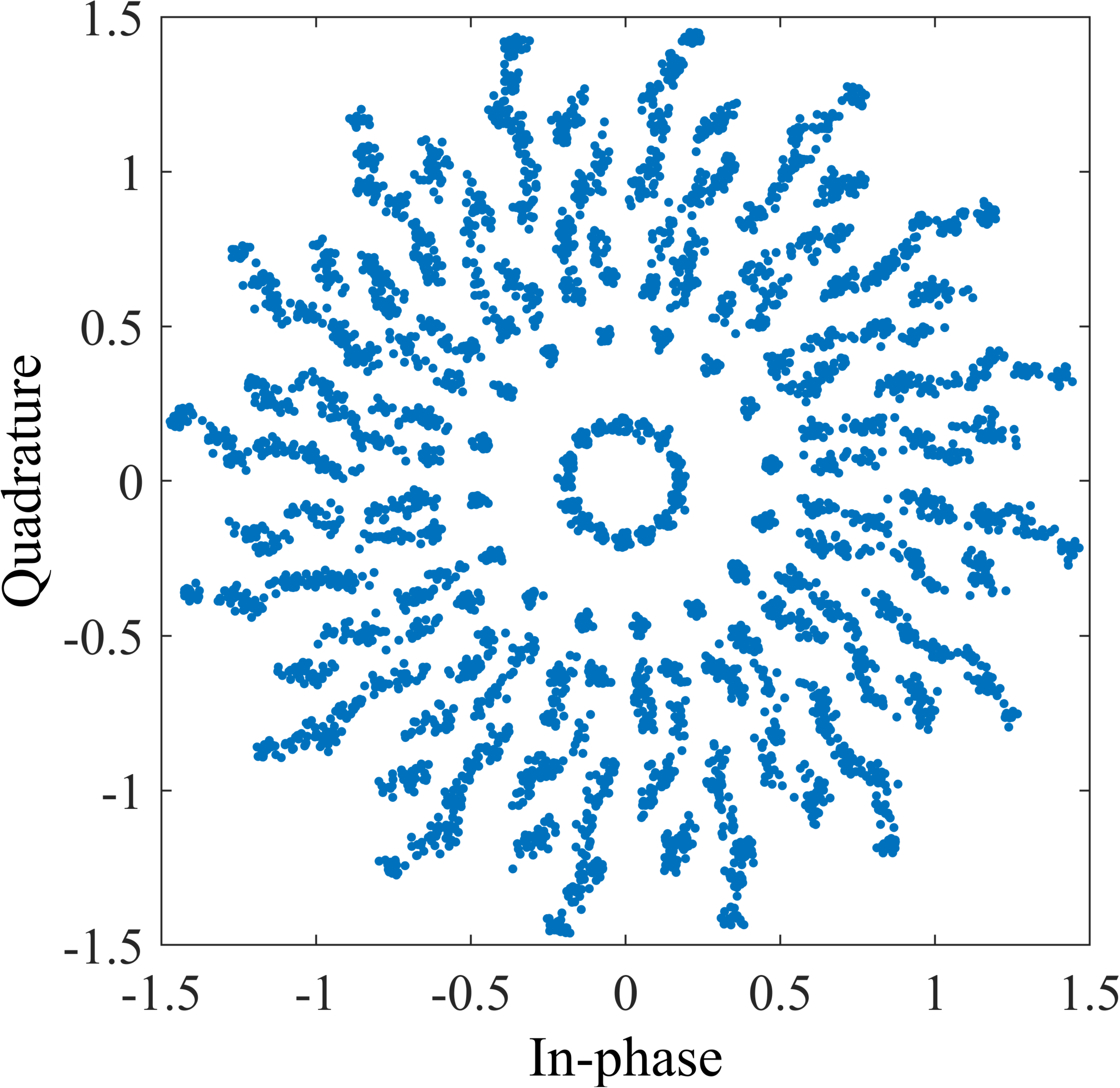} \label{prb2}}%
	\centering
	\subfloat[]{\includegraphics[width=0.48\linewidth]{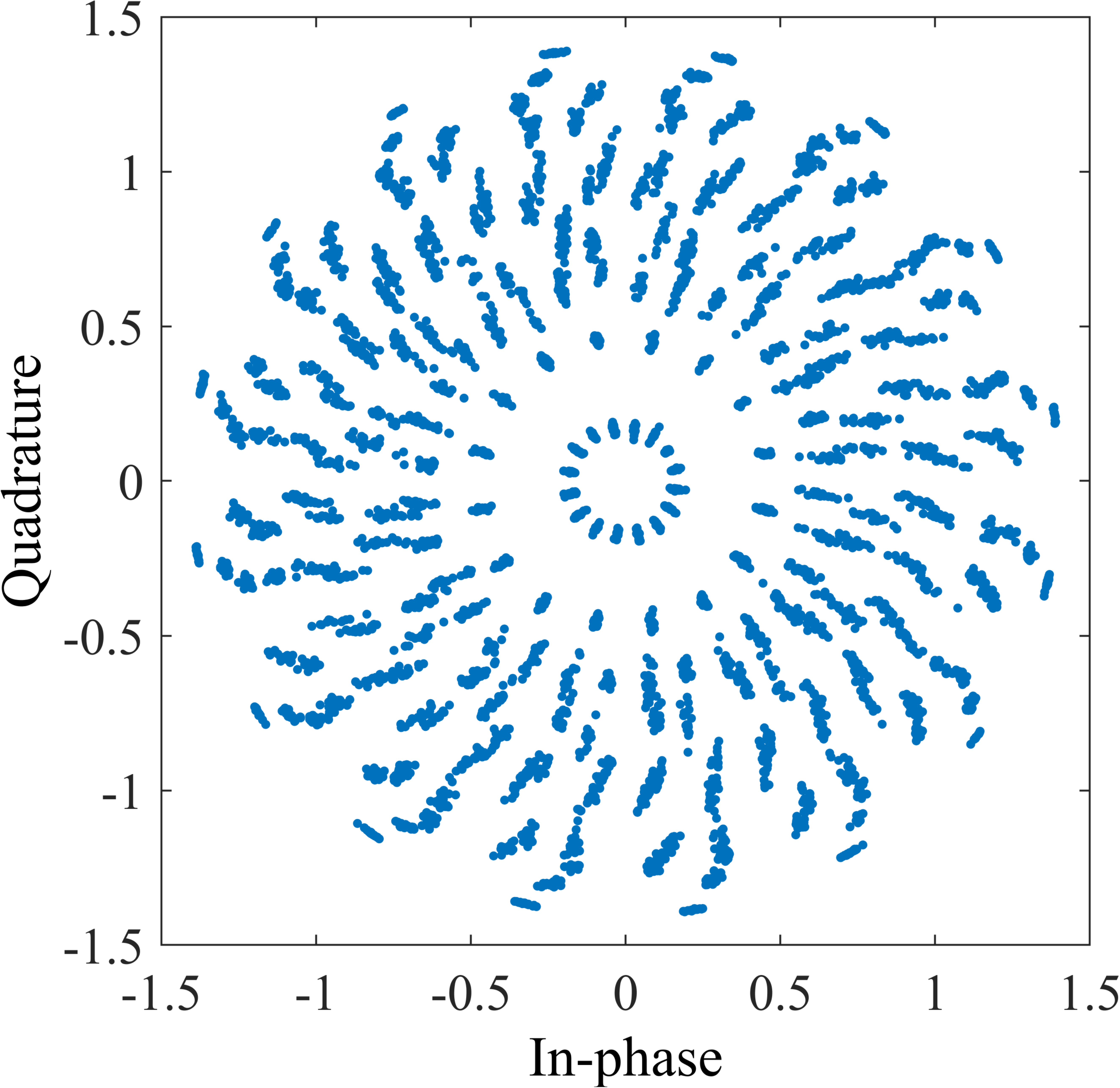} \label{proposed2}}%
	\caption{Measured constellations at 38 GHz. (a) Without signal reconstruction. (b) With baseline reconstruction. (c) With PC-enhanced baseline reconstruction. (d) With proposed reconstruction.}
	\label{fig:hardware_constellation2}
\end{figure}

The measured constellations at 5.4 GHz and 38 GHz for an MO of 16 are shown in Figs.~\ref{fig:hardware_constellation1} and \ref{fig:hardware_constellation2}, evaluated at IBOs of 4 dB and 6 dB, respectively. It can be observed that the unreconstructed received constellations exhibit severe PA-induced nonlinear distortion. The proposed algorithm achieves the best performance, with reconstructed symbols more tightly clustered around their ideal positions. The improvement is particularly pronounced at 38 GHz, where both the intra-cluster spread of individual symbols and the overall geometric structure of the constellation are significantly improved. At 5.4 GHz, the distortion is relatively mild, allowing the baseline algorithm to already perform well. As a result, the proposed method provides only marginal improvement in reducing intra-cluster symbol spread.

{Fig.~\ref{fig:hardware_results} presents the experimental SER results alongside the corresponding PAE values measured using \eqref{eq:PAE1}.} The proposed algorithm consistently outperforms both the baseline and the PC-enhanced baseline algorithms at 5.4 GHz and 38 GHz, with more pronounced gains at 38 GHz due to stronger nonlinear distortion. Specifically, it achieves approximately 0.2--0.5 dB of IBO reduction at 5.4 GHz and 1--2.5 dB of IBO reduction at 38 GHz compared to the baseline algorithm across  various MOs.
For a target SER of $10^{-3}$ and an MO of 64, the proposed algorithm provides PAE improvements of approximately 2.6\% at 5.4 GHz and 30.9\% at 38 GHz over the baseline. 
{It is noteworthy that the performance gains observed in simulations exceed those obtained experimentally. This discrepancy primarily arises because the PA model utilized in the simulations exhibits more severe nonlinearity than the actual hardware, naturally yielding larger algorithmic gains in the simulated environment. Furthermore, the idealized PA models cannot fully capture the intricate non-idealities of physical PAs, such as memory effects and thermal variations. Additionally, the experimental measurements are inevitably subject to practical imperfections, such as IQ imbalance, phase noise, and channel estimation errors, which fundamentally constrain the achievable performance.}
Despite these practical non-idealities, the proposed algorithm remains highly effective in achieving substantial IBO reduction and PAE improvement, demonstrating its robustness and practical applicability.

\begin{figure}[t]
	\centering
	\subfloat[]{\includegraphics[width=0.965\linewidth]{./figures/measured_5.4.pdf} \label{exp1}}%

	\centering
	\subfloat[]{\includegraphics[width=0.965\linewidth]{./figures/measured_38.pdf} \label{exp2}}
	\caption{Measured SER and PAE performance. (a) 5.4 GHz. (b) 38 GHz.}
	\label{fig:hardware_results}
\end{figure}

\section{Conclusion}
\label{sec:conclusions}
This paper investigated the  signal reconstruction problem in APTBM-based nonlinear communication systems by proposing a novel two-stage reconstruction framework. Specifically, the PA-induced nonlinear distortion was decoupled into a dominant component and a residual component. In the coarse reconstruction stage, the dominant amplitude and phase distortions were systematically mitigated by exploiting both the inherent structural properties of APTBM blocks and the pre-characterized PA nonlinearities. Subsequently, the fine reconstruction stage cast the residual distortion minimization as a nonconvex optimization problem, for which a  closed-form solution was derived. Comprehensive numerical simulations and  testbed experiments confirmed the superiority of the proposed framework. The results demonstrated that the proposed approach achieves substantial IBO reduction and remarkable PA efficiency enhancement while preserving transmission reliability.

{While the proposed two-stage reconstruction algorithm exhibits promising performance, the current study is  confined to single-antenna and single-carrier scenarios. Extending this framework to multiple-input multiple-output (MIMO) and OFDM systems represents a critical avenue for future work. In MIMO architectures, the spatial mixing of signals during beamforming disrupts the structural integrity of the APTBM constraints even prior to PA amplification. Furthermore, the spatial directivity of the generated beams is severely degraded by PA-induced nonlinearities, rendering the beamforming design and signal reconstruction intricately coupled. Analogous challenges arise in OFDM systems due to a fundamental domain mismatch: APTBM blocks are synthesized in the frequency domain, whereas  PA-induced nonlinear distortion occurs in the time domain. In MIMO-OFDM systems, the interplay between spatial beamforming and multi-carrier modulation further exacerbates these complications. To overcome these challenges, our future work will focus on developing a joint optimization framework that synergistically integrates beamforming, OFDM modulation, and signal reconstruction into a unified design paradigm.}

\bibliographystyle{IEEEtran}
\bibliography{./mypaper.bib}

\balance

\end{document}